\documentclass{emulateapj}
\usepackage{apjfonts}

\usepackage{graphics,graphicx,rotating}
\usepackage{natbib}
\usepackage{xspace}
\usepackage{amssymb}
\usepackage{amsmath}
\usepackage{epstopdf}
\usepackage{subfigure}
\usepackage{color}
\usepackage{soul}

 \usepackage{textcomp}
\usepackage{gensymb}

\shorttitle{Modeling the New Bullet Cluster, ZwCl008.8+52}
\shortauthors{Molnar and Broadhurst}

\newcommand{\simless} 
     {\ensuremath{\lower 3pt\hbox{$\rlap{\raise5pt\hbox{$\char'074$}}\mathchar"7218$}}}
\newcommand{\simgreat}
     {\ensuremath{\lower 3pt\hbox{$\rlap{\raise5pt\hbox{$\char'076$}}\mathchar"7218$}}}

\newcommand{\simgt}{\lower.5ex\hbox{$\; \buildrel > \over \sim \;$}}
\newcommand{\simlt}{\lower.5ex\hbox{$\; \buildrel < \over \sim \;$}}

\newcommand{\nop}{{\noindent}}
\newcommand{\etal}{{\it et al.}}

\newcommand{\LCDM}{{{\sc $\Lambda$CDM}}\xspace}

\newcommand{\HE}{hydrostatic equilibrium}

\newcommand{\DEG}{{$^{\circ}$}}

\newcommand{\MSUNFOUR}{{$10^{\,14}\,$\ensuremath{\mbox{\rm M}_{\odot}}}\xspace}
\newcommand{\TMSUNFOUR}{{$\times 10^{\,14}\,$\ensuremath{\mbox{\rm M}_{\odot}}}\xspace}

\newcommand{\KMSEC}{{$\rm km\;s^{-1}$}\xspace}

\newcommand*{\ltsim}{\ {\raise-.75ex\hbox{$\buildrel<\over\sim$}}\ }
\newcommand*{\gtsim}{\ {\raise-.75ex\hbox{$\buildrel>\over\simsaus$}}\ }
\newcommand*{\proptosim}{\ {\raise-.75ex\hbox{$\buildrel\propto\over\sim$}}\ }

\newcommand{\MACH}{{\cal M}\xspace}

\newcommand*{\CHANDRA}{\emph{Chandra}\xspace}

\newcommand*{\FLASH}{\emph{FLASH}\xspace}

\newcommand*{\ZWCLEIGHT}{ZwCl008\xspace}

\newcommand*{\CIZSAUSAG}{CIZA J2242.8+5301\xspace}

\begin{document}

\title{Multi-Phenomena Modeling of the New Bullet Cluster, ZwCl008.8+52, \\
                       using $N$-body/\-hydro\-dynamical Simulations}

\author{
S. M. Molnar\altaffilmark{1} and T. Broadhurst\altaffilmark{2,3}
}

\altaffiltext{1}{Institute of Astronomy and Astrophysics, Academia Sinica, P. O. Box 23-141,
                      Taipei 10617, Taiwan}

\altaffiltext{2}{Department of Theoretical Physics, University of the Basque Country, Bilbao 48080, Spain}
                                            
\altaffiltext{3}{Ikerbasque, Basque Foundation for Science, Alameda Urquijo, 36-5 Plaza Bizkaia 48011, Bilbao, Spain}


\begin{abstract}
We use hydrodynamical/$N$-body simulations to interpret the newly 
discovered Bullet-cluster-like merging cluster, ZwCl 0008.8+5215 (\ZWCLEIGHT hereafter), 
where a dramatic collision is apparent from multi-wavelength observations. 
We have been able to find a self-consistent solution for the radio, X-ray, 
and lensing phenomena by projecting an off-axis, 
binary cluster encounter viewed just after first core passage. 
A pair radio relics traces well the leading and trailing shock fronts that our 
simulation predict, providing constraints on the collision parameters. 
We can also account for the observed distinctive comet-like X-ray 
morphology and the positions of the X-ray peaks relative to the two lensing mass 
centroids and the two shock front locations.
Relative to the Bullet cluster, the total mass is about 70\% lower, 
$1.2\pm0.1\times 10^{15} \,M_\odot$, with a correspondingly lower infall velocity,
$1800 \pm 300\,$\KMSEC, and an impact parameter of $400 \pm 100$ kpc. 
As a result, the gas component of the infalling cluster is not trailing significantly 
behind the associated dark matter as in the case of the Bullet cluster.
The degree of agreement we find between all the observables provides 
strong evidence that dark matter is effectively collisionless on large scales
calling into question other claims and theories that advocate modified gravity.
\end{abstract}

\section{Introduction}
\label{S:Intro}

The Bullet cluster has provided particularly direct evidence for the existence
of dark matter by displaying a large offset between the 
the gas component, marked by X-ray emission,
and dark matter traced by gravitationally lensed images
\citep{MarkevitchET02}. 
This merging cluster also demonstrates that dark matter is collisionless to high precision, 
because, although the X-ray shocks unambiguously reveal that two clusters 
have clearly just collided (Markevich 2005), the two dark mater centroids are still 
centered on their respective galaxy members  \citep{CloweET2006}. 
When two clusters suffer gravitational encounters their respective member galaxies being 
relatively small are unlikely to collide with each other. 
Therefore the cluster member galaxies are effectively 
collisionless, hence, If the dark matter has an associated collisional cross section 
it may be revealed by a relative displacement between the two dark matter lensing 
centroids and their respective member galaxy distributions. 
The amplitude of this displacement is proportional to the self-interaction 
cross section of dark matter particles.
Using this property, the Bullet cluster was also the first to be used to put upper limits 
on the self-interaction cross section of dark matter particles 
providing quantitative evidence that the dark matter is 
collisoinless \citep{MarkevitchET2004ApJ606}.

Since the discovery of the Bullet cluster, more merging clusters have been found 
with offset between the centroids of the intra-cluster gas and the dark matter 
(MACS J1149.5+2223: \citealt{GolovichET2016ApJ831} ; 
CL0152-1357: \citealt{MolnarET2012ApJ748}
DLSCL J0916.2+2951: \citealt{DawsonET2012}; 
ACT-CL J01024915: \citealt{MenanteauET2012}; 
MACS J0717.5+3745: (Mroczkowski et al. 2012)   \citealt{MroczkowskiET2012}; 
A1758N: \citealt{RagozzineET2011}; 
A2744: \citealt{MertenET2011}
A2163: \citealt{OkabeET2011}; 
ZwCL0008.8+5215: \citealt{WeerenET2011AA528}; 
CL0152-1357: \citealt{MassardiET2010}; 
A1240: \citealt{BarrenaET2009}; 
MACS J0025.4-1222: \citealt{BradacET2008}; 
A520: \citealt{MahdaviET2007};
Bullet Cluster: \citealt{CloweET2004})

The collisions of massive clusters of galaxies provide a unique possibility
to study the nature of dark matter as these collisions are the most energetic 
of all phenomena (after the Big Bang)
with the power to separate the collisional gas from the confining dark matter 
gravitational potential, including the galaxies, 
which provide an effectively collisionless tracer population.
Any significant inconsistencies in accounting for the relative dynamics of 
the components would therefore be of fundamental importance. 
To examine this possibility to the fullest requires, of course, appropriate simulations 
that incorporate dark matter and gas. The first test to make in this respect is to 
compare the best data with collisionless dark matter, 
which is well described by $N$-body simulations together with
an appropriate hydrodynamical numerical scheme for the associated hot gas. 
The gas is much harder to model, but several high quality codes have been constructed 
and tested for this purpose with different levels of approximation and numerical schemes 
 -- ranging from adaptive mesh refinement (AMR) to smooth particle hydrodynamics  (SPH) --
and applied to the Bullet cluster \citep{LageFarrar2014ApJ787,MastBurk08MNRAS389p967,SpringelFarrar2007MNRAS380};
the Sausage cluster (\CIZSAUSAG; \citealt{MolnarBroadhurst2017,Donnert2017MNRAS471})
El Gordo (ACT-CT J0102-4915; \citealt{ZhangET2015ApJ813,MolnarBroadhurst2015});
A1758N \citep{MachadoET2015};
A1750 \citep{Molnar2013ApJ779}; and 
CL0152-1357 \citep{MolnarET2012ApJ748}; 
for a review see \citealt{Molnar2015Frontiers}.

Several examples of extreme collisions have now been analyzed in detail for which the speed of
the infalling cluster exceeds the sound speed of the gas, thus merging shocks are generated. 
These shock fronts, in some systems, are traced by radio ``relics'', large-scale diffuse 
synchrotron emission (\citealt{Ensslin1999}; for a recent review see \citealt{FerettiET2012}). 
There are also very bright gas features corresponding compressed gas that, although subsonic,
are relatively dense so that the X-ray emission is enhanced showing a large scale
cometary morphology, such as El Gordo \citep{MenanteauET2012}.
This morphology has been readily explained as an off axis binary collision in
self-consistent $N$-body/\-hydro\-dynamical numerical simulations assuming 
zero self-interacting cross section for the dark matter
\citep{ZhangET2015ApJ813,MolnarBroadhurst2015}.
Radio relics in El Gordo delineate a bow shock and a back shock.
We define bow shock as the shock front in the gas of the main cluster 
generated by the infalling cluster, which resembles a bow shock of a bullet 
flying through air.
The back shock is the shock front propagating in the 
gas of the infalling cluster in the opposite direction to the cluster.
The back shock was detected by X-ray and SZ observations 
\citep{BotteonET2016,BasuET2016AA591}.
Another such example is the Sausage cluster (\CIZSAUSAG), in which, in addition
to tidally compressed gas between the two merging clusters highlighted by X-ray
emission, there are impressive radio ``relics'' marking the location of the bow shock front 
in particular with evidence of a back shock in the radio observations 
\citep{MolnarBroadhurst2017,WeerenET2010Sci}.

These simulations are demanding in time and analysis despite the inherent simplicity 
of binary encounters, because the gas interactions must be resolved well spatially and 
temporally, and there is a wide rage of masses and impact parameters to explore and 
projection angles involved. 
With some intuition in exploring the output of these models and the improvements in data quality,
we can become more efficient by narrowing the ranges of the relevant initial conditions.

This numerical work has, in the above cases, found consistency with the collisionless dark matter 
hypothesis by finding compelling agreement among the independent observables
including the lensing, radio relic, X-ray and SZ data that are increasingly being obtained for 
examining the binary collision clusters. 
A discrepancy that has been advanced for the ``Pandora''  cluster (A2744) in terms of dark matter 
offsets is not so readily modeled as the system this multi-model in complexity 
\citep{LamET2014ApJ797,Zitrin2014ApJ793,MertenET2011},
so such statements are qualitative at present.
Further clarification would also benefit from a deeper weak lensing analysis \citep{MedezinskiET2016ApJ817}.
Even for binary collisions it is very clear that approximate or misleading eyeball statements regarding 
the basic collision parameters are often easily cleared up uniquely, 
thanks to comparison with the numerical calculations.
In particular, statements regarding the relative impact velocity and the time after first core passage 
at which the system is being viewed are often highly
uncertain without the guidance of a full self-consistent simulation 
\citep[e.g.][]{GolovichET2016ApJ831,NgET2015}.

The issue of the relative velocity is particularly interesting as it
provides a definitive test of the \LCDM scenario.
The distribution of dark matter halo relative velocities has been calculated in detail 
by several $N$-body simulations based on \LCDM 
(e.g., \citealt{BouillotET2015,ThompsonNagamine2012}), 
as well as including approximate hydro interactions \citep{CaiET2009}.
The first such consistency test was inspired by the Bullet cluster. 
The infall velocities in the Bullet cluster, $\sim 3000\,$\KMSEC, 
derived using detailed $N$-body/hy\-dro\-dy\-nam\-ical simulations based on 
multifrequency observations seemed to be too high for \LCDM models 
\citep{MastBurk08MNRAS389p967,SpringelFarrar2007MNRAS380}.
Analyzing large cosmological numerical simulations, 
\cite{LeeKomatsu2010ApJ718} and \cite{ThompsonNagamine2012}
found that the Bullet cluster is incompatible with \LCDM\ models. 
\cite{BouillotET2015} arrived at the same conclusion adopting a new halo 
finder algorithm, ROCKSTAR, and using extreme value statistics. 
In contrast, \cite{WatsonET2014} and \cite{LageFarrar2015JCAP} 
using different cosmological simulations concluded that the Bullet cluster is 
not excluded by the \LCDM.
\cite{KraljicSarkar2015}, 
using the ``Dark Sky Simulations'', the ROCKSTAR algorithm, 
and extreme value statistics, found that the number of Bullet-cluster-like 
massive halos is $\sim\,$0.1, i.e., the Bullet cluster is compatible with the \LCDM models.
However, more high-infall velocity merging clusters have been identified recently
(Abell 2744: \citealt{OwersET2011}; 
CL J0152-1347: \citealt{MolnarET2012ApJ748};
MACS J0717.5+3745: \citealt{MaET2009} and \citealt{SayersET2013};
El Gordo: \citealt{MolnarBroadhurst2015}).

The probability to find all of these massive systems simultaneously 
based on cosmological simulations of \LCDM models has not been assessed yet.
It is an open question today wether their infall velocities are compatible with
the predictions of our standard \LCDM models. 
Clearly, a statistical sample of such clusters is of great interest to clarify this question further
A confirmation of a sample of merging clusters with high infall velocities 
could be a serious challenge to the standard \LCDM models
\cite[e.g.,][]{Molnar2015,KraljicSarkar2015}.

Here we apply our well-tested \FLASH based 3-dimensional (3D) Hydro/$N$-body code to a recently 
discovered Bullet-cluster-like binary collision encounter that has been recently 
recognized \citep{GolovichET2016ApJ831}, 
but for which self-consistent simulations have yet not been applied. 
This system has the advantage of showing a clear cut pair of radio relics, 
a distinct bullet like X-ray morphology, 
weak gravitation lensing based mass centroids for the two interacting components, and 
line of sight redshift information for cluster member galaxies \citep{GolovichET2016ApJ831}.
We follow the time evolution of the merging shocks until they run out of the intracluster gas,
and test the assumption of collisionless dark matter inherent to the standard LCDM cosmology.
We make use of the AMR code, \FLASH allowing us to follow 
the shocks in the low density intracluster gas, that is not well represented in the fixed grid 
Eulerian scheme and codes based on SPH 
(for a comparison between AMR and SPH simulations see, 
e.g., \citealt{MitchellET2009,AgertzET2007}).

The structure of this paper is as follows.
In Section~\ref{S:ZWCL0008} we summarize results
from previous analyses of \ZWCLEIGHT based on multifrequency
observations and numerical simulations.
We describe our simulation setup for modeling of \ZWCLEIGHT
as a binary merger in Section~\ref{S:SIMULATIONS}.
Section~\ref{S:RESULTS} presents our results, a dynamical model for 
\ZWCLEIGHT, a discussion on the dynamics of merging shocks 
in clusters similar to \ZWCLEIGHT, and a comparison with the Bullet cluster.
Section~\ref{S:CONCLUSIONS} contains our conclusions.
We adopt a spatially flat \LCDM cosmology with h = 0.7, 
$\Omega_m = 0.3$, thus $\Omega_\Lambda = 0.7$.
Unless stated otherwise, the quoted errors represent 68\% Confidence levels (CLs).

\section{\ZWCLEIGHT: The Newly Discovered Bullet-like Cluster}
\label{S:ZWCL0008}

The merging cluster  \ZWCLEIGHT, at a redshift of 0.1032, 
was observed by \cite{WeerenET2011AA528} using 
the Giant-Meterwave Radio Telescope (GMRT) 
at 241 and 640 MHz and the Westerbrook Synthesis Radio Telescope
(WSRT) at 1.3-1.7 GHz. 
They found two radio relics to the east and west from 
the X-ray peak emission, with the eastern relic much more elongated
(first panel in Figure~\ref{F:OBSMOD}).
The spectral indices of both relics are steepening towards
the cluster center, suggesting that they are moving outward, 
away from the center of the merging system.
The spectral indices at the front of the east and west relics 
were reported to be $-1.2\pm0.2$ and $-1.0\pm0.15$.
Adopting these as the spectral indices of the injection distribution,
they derive Mach numbers $\MACH = 2.2_{-0.1}^{+0.2}$ and $\MACH = 2.4_{-0.2}^{+0.4}$
for the east and west relics; the polarizations were constrained to 5\%-25\% and 5\%-10\%.

More recently \cite{KierdorfET2017AA} carried out 
high-frequency radio observations of \ZWCLEIGHT 
at 4.85 and 8.35 GHz with the Effelsberg telescope.
They found a polarization fraction of the eastern relic 
between 20\% and 30\%, 
and derived a Mach number of $\MACH = 2.35\pm0.1$, 
in agreement with previous radio measurements of 
\cite{WeerenET2011AA528}.

Most recently, \cite{GolovichET2017} carried out a dynamical 
analysis of \ZWCLEIGHT based on detailed radio (JVLA), 
optical (HST, Subaru/SuprimeCam, Keck/DEIMOS)
and X-ray (Chandra/ACIS-I) observations and weak lensing 
observations (Subaru/HST) to estimate masses of 
M$_{200,1} = 5.73_{-1.81}^{+2.75}\,$ \TMSUNFOUR 
and M$_{200,2} = 1.21_{-0.63}^{+1.43}\,$\TMSUNFOUR
for the main and infalling cluster respectively, 
which is a mass ratio of $\simeq 5$.

\cite{GolovichET2017} used this information as input in their 
dynamical model, which is based on fixed NFW 
\citep{NFW1997ApJ490p493} gravitational potentials
for the two components and zero impact parameter, ignoring
the effects of gas, gravitational tidal effects, mass loss,
and integrating the equations of motion numerically.
Golovich et al. estimated the merger velocity at pericenter, 
$V_p$, and obtained $V_p = 1800^{+400}_{-300}$ \KMSEC.
The inclination angle relative to the plane of the sky, $\theta$, 
was constrained to $6.6\degree \simless\; \theta \simless\; 31\degree$,
which is consistent with the direct constraint derived from radio polarization 
measurements: $\theta\; \simless\; 40\degree$.
Golovich et al. concluded that the gas of the two merging subclusters
is still moving outward, and derived the phase of the system as 
either 0.76$_{-0.27}^{+0.24}$ Gyr or 1.3$_{-0.35}^{+0.90}$ Gyr after the
first core passage for the outgoing phase and infalling phase 
(after the turnover) respectively.
They could not distinguish between the outgoing and returning phase
because their model is time symmetric, 
includes only the dark matter and not gas, hence the X-ray emission,
which provides information of the gas, could not be not interpreted.

\section{Modeling \ZWCLEIGHT using hydrodynamical simulations}
\label{S:SIMULATIONS}

Our main goals were to obtain a reasonable physical model for
the newly discovered Bullet-cluster-like merging cluster, \ZWCLEIGHT,
using $N$-body/\-hydro\-dynamical simulations, and thus estimate the infall velocity 
and constrain the phase of the collision with high precision.
We have not carried out a systematic search for all the initial parameters
and determined their errors with statistical measures, 
which would require many more simulations.
The errors we quote for the results from our simulations are conservatively estimated.

\subsection{Details of the simulations}
\label{SS:ICOND}

We modeled \ZWCLEIGHT in 3D using an Eulerian $N$-body/hydrodynamic 
code \FLASH (developed at the Center for Astrophysical Thermonuclear Flashes
at the University of Chicago; \citealt{Fryxell2000ApJS131p273,Ricker2008ApJS176}).  
\FLASH is a publicly available AMR code, which can be run in parallel computer architectures.
We assumed a binary merger for \ZWCLEIGHT and 
included dark matter and gas self-consistently taking their gravity into account dynamically.
We used our well-established method to carry out merging 
cluster simulations 
\citep{MolnarBroadhurst2017,MolnarBroadhurst2015,Molnar2013ApJ779,Molnar2013ApJ774,MolnarET2012ApJ748}.
For our simulations, we adopted a large box size (13.3 Mpc on a side) 
to capture the outgoing merger shocks and avoid loosing mass 
during the time we ran our simulations.
Our highest resolution, 12.7 kpc was reached at the cluster centers, merger shocks, 
and in the turbulent regions behind the shocks.
We chose 3D Cartesian coordinate system, $x,y,z$, with the $x,y$ plain containing the 
centers of the clusters and the initial (relative) velocity vector of the infalling cluster
in the positive $x$ direction.
We included shock heating, the most important non-adiabatic process in merging clusters,
and ignored other heating and cooling processes.

%
%
\begin{deluxetable}{lcccccc}[t]
\tablecolumns{11}
\tablecaption{                       \label{T:TABLE1} 
 IDs and input parameters for different models used in our hydrodynamical simulations.\\
} 
\tablewidth{0pt} 
\tablehead{ 
 \multicolumn{1}{l}                {ID\,\tablenotemark{a}}       &
 \multicolumn{1}{c}   {M$_{vir1}$\,\tablenotemark{b}}     &
 \multicolumn{1}{c}   {c$_{vir1}$\tablenotemark{b}}        &
 \multicolumn{1}{c}   {M$_{vir2}$\,\tablenotemark{c}}          &
 \multicolumn{1}{c}   {c$_{vir2}$\tablenotemark{c}}        &
 \multicolumn{1}{c}              {P\,\tablenotemark{d}}          &
 \multicolumn{1}{c}   {V$_{in}$\,\tablenotemark{e}}       
 }
 \startdata  
   P400V18B       &   7.0    &   6     &   5.0   &  8  &  400  &  1800   \\ \hline
   P100V15G      &   6.0    &   6     &   1.5   &  8  &  100  &  1500   \\ \hline
   P300V18         &   7.0    &   6     &   5.0   &  8  &  300  &  1800   \\ \hline
   P500V18         &   7.0    &   6     &   5.0   &  8  &  500  &  1800   \\ \hline
   P400V18M1    &   7.0    &   6     &   5.5   &  8  &  400  &  1800   \\ \hline
   P400V18M2    &   7.0    &   6     &   4.5   &  8  &  400  &  1800   \\ \hline
   P400V18M3    &   7.5    &   6     &   5.5   &  8  &  400  &  1800   \\ \hline
   P400V18M4    &   7.5    &   6     &   5.0   &  8  &  400  &  1800   \\ \hline
   P400V18M5    &   7.5    &   6     &   4.5   &  8  &  400  &  1800   \\ \hline
   P400V18M6    &   6.5    &   6     &   5.5   &  8  &  400  &  1800   \\ \hline
   P400V18M7    &   6.5    &   6     &   5.0   &  8  &  400  &  1800   \\ \hline
   P400V18M8    &   6.5    &   6     &   4.5   &  8  &  400  &  1800   \\ \hline
   P400V15         &   7.0    &   6     &   5.0   &  8  &  400  &  1500   \\ \hline
   P400V20         &   7.0    &   6     &   5.0   &  8  &  400  &  2000   
 \enddata
\tablenotetext{a}{IDs of the runs indicate the impact parameters in kpc and the infalling velocities in 
                            in units of 100 \KMSEC}
\tablenotetext{b}{Virial mass  in \MSUNFOUR and concentration parameter for the main cluster (1).}
\tablenotetext{c}{Virial mass in \MSUNFOUR and concentration parameter for the infalling cluster (2).}
\tablenotetext{d}{Impact parameter in units of kpc.}
\tablenotetext{e}{Infall velocity of cluster 2 in \KMSEC.
\vspace{.4 cm}}
\end{deluxetable}  

%
%
\begin{figure*}[t]
\includegraphics[width=.33\textwidth]{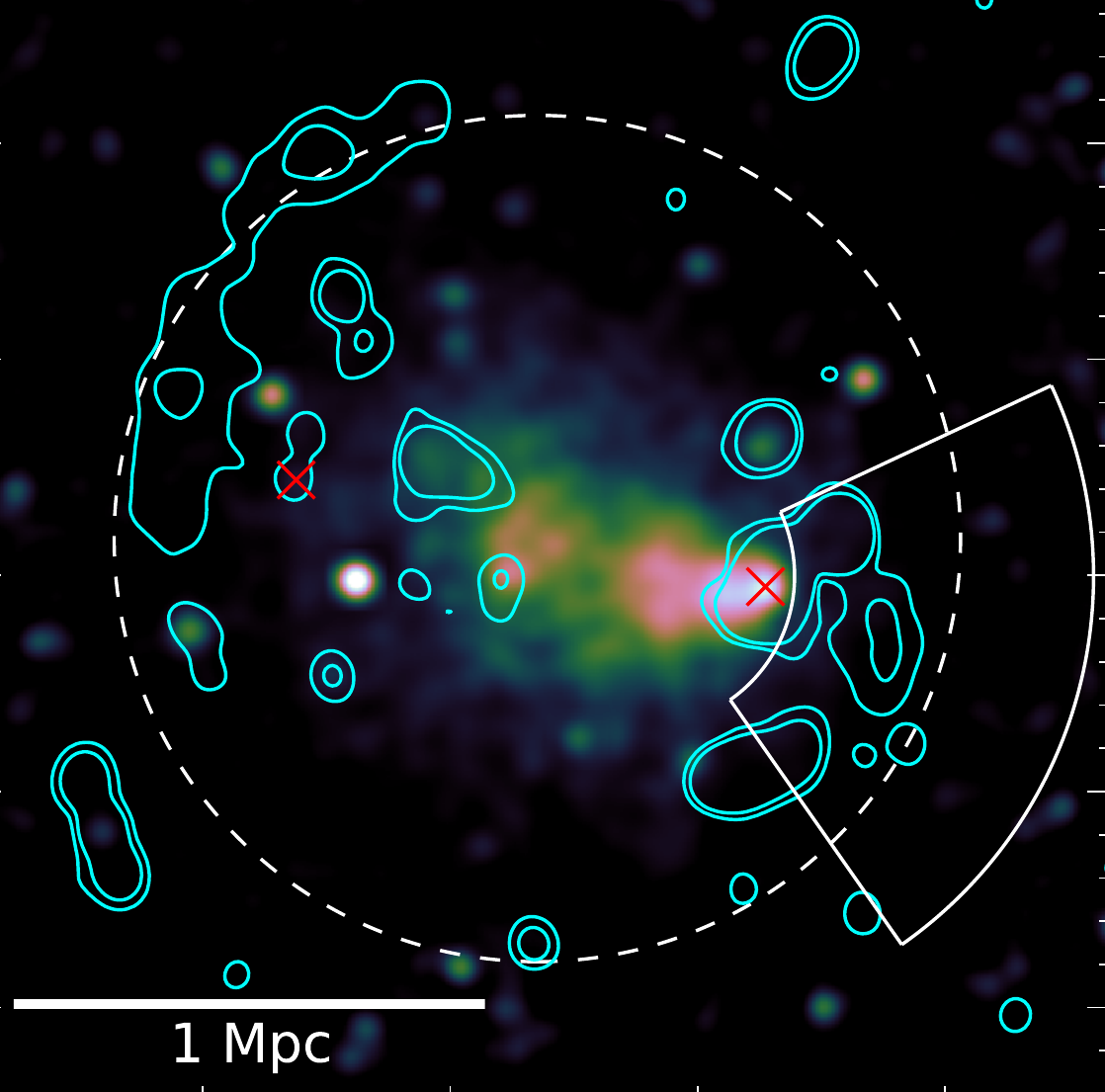}  
\includegraphics[width=.32\textwidth]{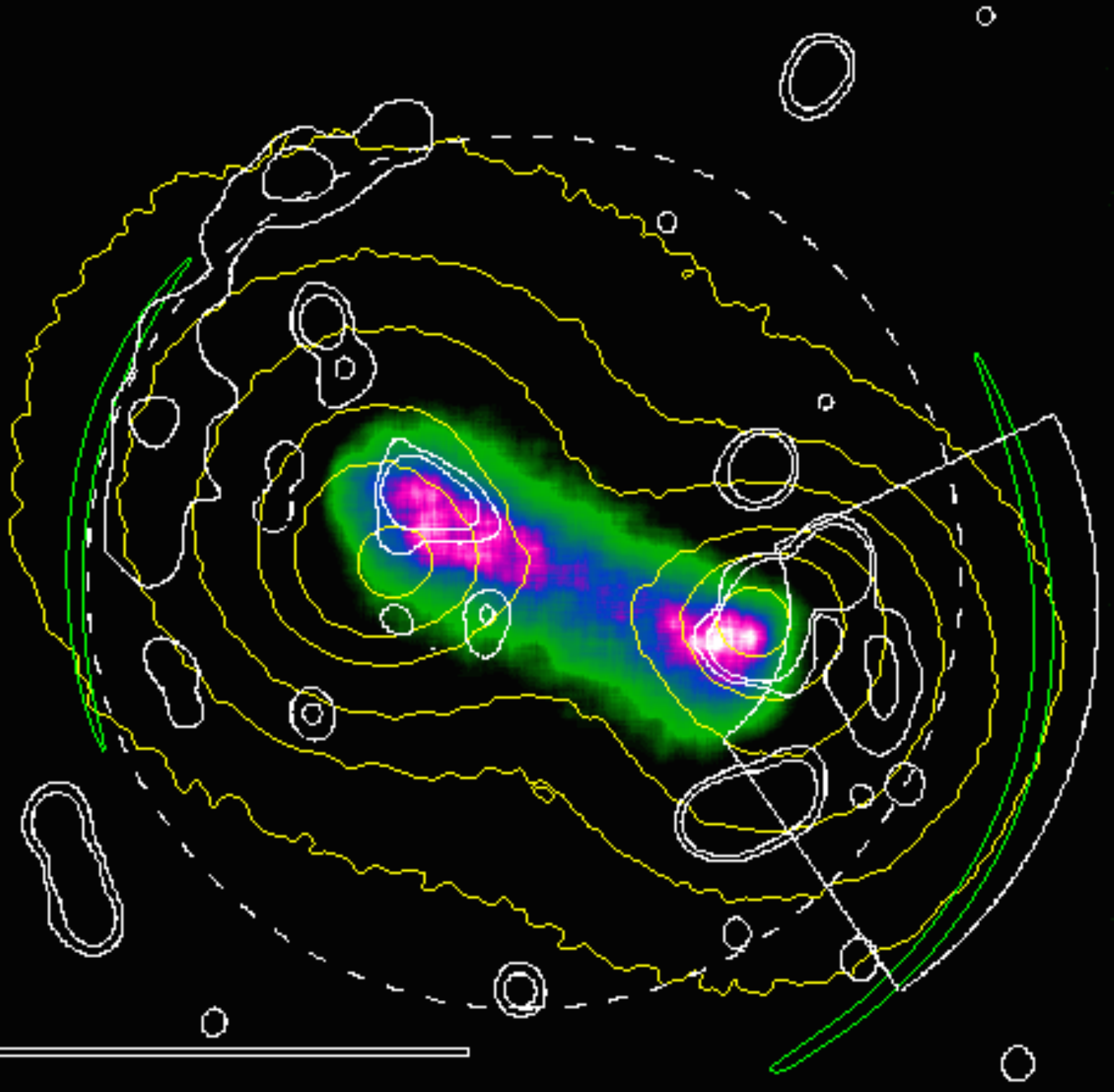}  
\includegraphics[width=.33\textwidth]{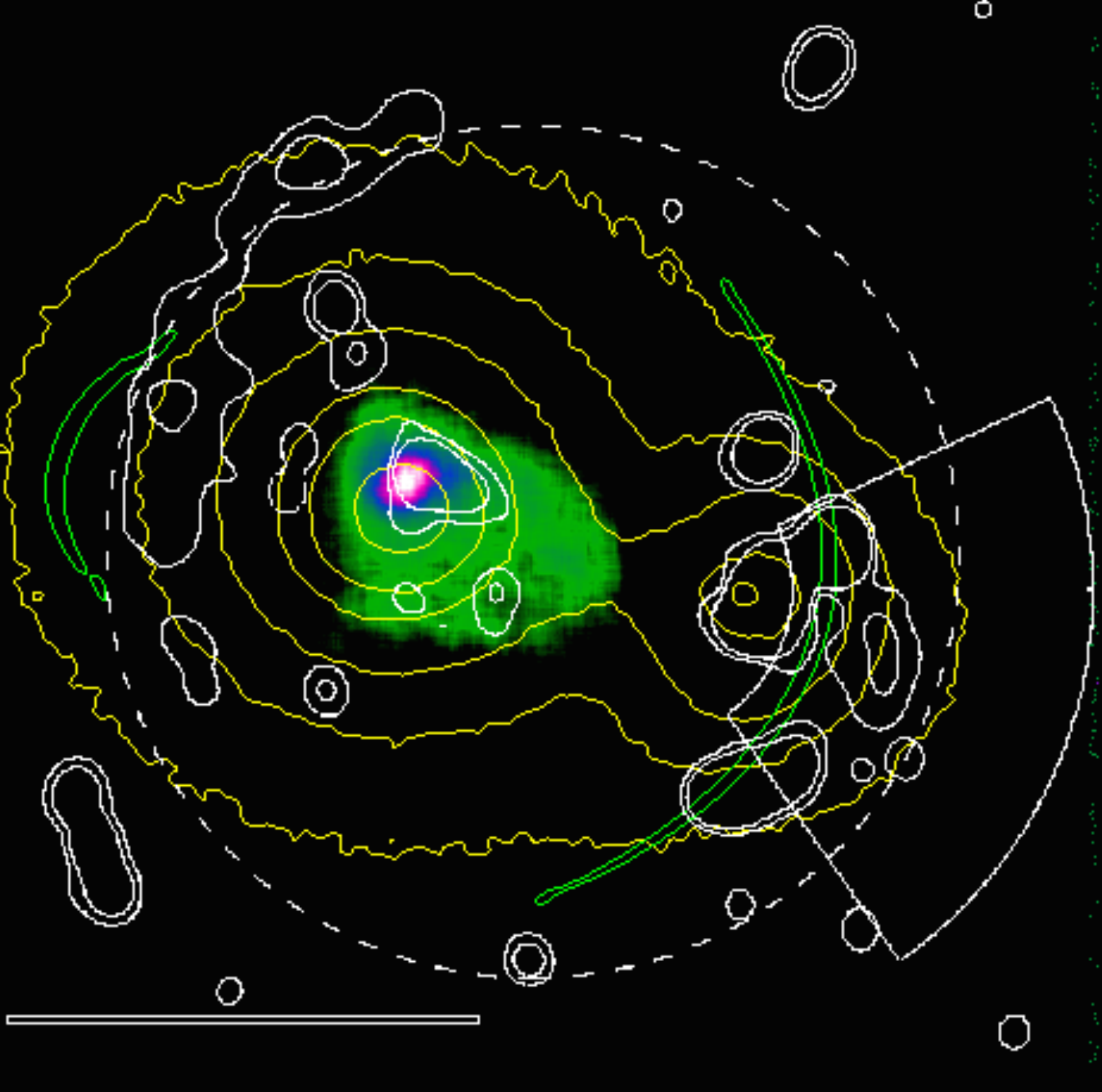} 
\caption{
1st panel: \CHANDRA X-ray color image of \ZWCLEIGHT with the
WSRT radio radio contours (cyan) based on WSRT observations 
overlaid (from \citealt{GolovichET2017,WeerenET2011AA528}). 
The white dashed circle and the horizontal bar represent R$_{500}$ and a 
length of 1 Mpc. The white annular sector marks the proposed shock
region (see \citealt{GolovichET2017}).
The two BCGs are shown by red crosses.
2nd panel: On the same scale, 
simulated X-ray image based on our best model (run P400V18B)
at t = 428 Myr after the 1st core passage 
with the contours of the dark matter distribution overlaid (yellow).
The green contours represent the outgoing shocks.
The viewing angle was chosen to match the observations (see text for details).
The bow shock on the right ahead of the infalling cluster moving to the right,
the back shock on the left is moving to the left, to the opposite direction
to the motion of the infalling cluster.
3rd panel: On the same scale, our 
run P10V15G with initial conditions suggested by \cite{GolovichET2017}. 
There is only one X-ray peak, which is associated with the gas of the main cluster.
The gas of the infalling cluster has been stripped off as a result of the relatively large  
velocity of the infalling cluster and the small impact parameter.
\vspace{0.3 cm}
\label{F:OBSMOD}
}
\end{figure*} 
\vspace{1.3 cm}

The initial models of the clusters were assumed to have spherical geometry 
with cut offs of the dark matter and gas density at the virial radius, $R_{\rm vir}$.
We assumed an NFW model \citep{NFW1997ApJ490p493} for the dark 
matter distribution, 
\begin{equation} \label{E:NFW}
      \rho_{DM} (r) =  { \rho_s  \over x(r) (1 + x(r))^2}; \,\,r \le R_{\rm vir}
,
\end{equation}
\nop
where $x(r) = r/r_s$ ($r_s = R_{vir}/c_{vir}$) and $\rho_s$, are scaling parameters 
for the radius and the amplitude of the density, and $c_{vir}$ is the concentration parameter.
We adopted a non-isothermal $\beta$ model for the gas density, 
\begin{equation}  \label{E:BETAMODEL}
      \rho_{gas}(r) =  { \rho_0  \over (1 + y^2)^{3 \beta /2} }; \,\,r \le R_{\rm vir}
,
\end{equation}
\nop 
where $y = r/r_{core}$, is the scaling parameters for the radius, $r$ 
(in units of the core radius,  $r_{core}$), and 
$\rho_0$, is the density at the center of the cluster.
The exponent, $\beta$ determines the fall off of the density 
distribution at large radii. We adopted $\beta=1$, suggested by 
cosmological numerical simulations for the large scale distribution 
of the intracluster gas in equilibrium (excluding the filaments; 
see \citealt{Molnet10ApJ723p1272}).

We derived the gas temperature as a function of the radius, $T(r)$,
assuming \HE\ adopting $\gamma = 5/3$ for the ideal gas equation of state.
We used a gas fraction, $f_{gas}$,  $f_{gas} = 0.14$, and represented baryons in galaxies 
together with the collisionless dark matter particles, 
since, for our purposes, galaxies can also be considered collisionless.

It is less straightforward to model a stable dark matter density distribution 
than a gas distribution. The hydrostatic equilibrium assumption provides
a stable distribution for the gas, but the dark matter, modeled as particles, 
has no pressure, interacting only gravitationally with itself and the gas, 
which means that they move on orbits in the potential of the cluster. 
We use the local Maxwellian approximation for the amplitude 
of the velocities of the dark matter particles: we randomly sample 
a Maxwellian distribution with a velocity dispersion as
a function of radius, $r$, $\sigma_v (r)$ derived from 
the Jeans equation assuming that the distribution of $\sigma_v (r)$ is 
isotropic \citep{LokasMamon2001MNRAS321}.
We assumed an isotropic distribution for the direction of the velocity vectors 
(for more details of the set up for our simulations see \citealt{MolnarET2012ApJ748}).

\subsection{\FLASH\ Runs}
\label{SS:RUNS}

We have run a series of \FLASH simulations varying the 
initial masses, concentration parameters, impact parameter, 
and infall velocity of our models.
Our aim was to find a physical model for \ZWCLEIGHT with 
a reasonable agreement with observations.
Our simulations were constrained by the masses and positions 
of the dark matter centers derived from weak gravitational lensing, 
X-ray morphology \citep{GolovichET2017}, 
and the positions of the outgoing merging shocks inferred from radio 
observations \citep{WeerenET2011AA528}.
The long bright radio relic to the east most likely marks the location of the 
back shock, as in the \CIZSAUSAG cluster, as demonstrated by \citealt{MolnarBroadhurst2017}), 
due to the limb brightening of the spherical surface of the relic viewed in projection 
The position of the bow shock associated with the infalling 
cluster is much less certain. This forward shock is not detected by the
X-ray observations \citep{GolovichET2017}).
We estimate from our models that it should lie significantly beyond the small radio 
relic that we nevertheless do associate with this shock.
The reason for this displacement can be simply geometric in origin,
because such shock surfaces are convex in shape, thus, in projection, radio
relics on their surface may often appear to lie behind the front when viewed from the side 
despite being generated by the shock (e.g, planetary nebulae appear like a ring,
but they are spherical shells). In other words, because radio relics cannot be expected
to trace the full shock surface uniformly, but have a patchy covering, thus it is 
likely to see radio relics appearing ``inside'' the projected shock front, and only occasionally 
marking the projected shock itself when a relic happens to cover some of the projected area 
of the shock front. When that happens, the radio relic can be of high surface brightness
as the projected radio emission (which is optically thin) adds up in projection.
Indeed, notable examples of large shock fronts of anomalously bright such are known, 
in particular, the bow shock of \CIZSAUSAG \citep{WeerenET2011MNRAS418}
that modeling has been shown to coincide with their observed projected shock front 
\citep{MolnarBroadhurst2017}.

We have carried out a suit of simulations to provide a rough 
estimate on the errors on the infall velocity, impact parameter, and masses
of the merging system before the collision. 
A systematic parameter search is currently beyond reach of
conventional high speed computing resources based on CPUs.

Table~\ref{T:TABLE1} contains a list of initial parameters 
of those simulations we discuss in this paper. 
This is a narrow subset of parameter space that contains our best solutions 
with enough spread to illustrate the observable effects of moving away 
from the best-fit solution.
The first column contains the IDs of our runs as 
$PijkVmn$, where $Pijk$ is the impact parameter in kpc, 
and $Vmn$ is the infall velocity in units of 100 \KMSEC.
In columns 2 to 5 we show the the virial masses (in units of \MSUNFOUR) 
and concentration parameters ($c_{vir}$) of the two subclusters.
Columns 6 and 7 contain the impact parameters, $P$,  
and infall velocities ($V_{in}$) in \KMSEC.

%
%
\begin{figure*}[t]
\includegraphics[width=.33\textwidth]{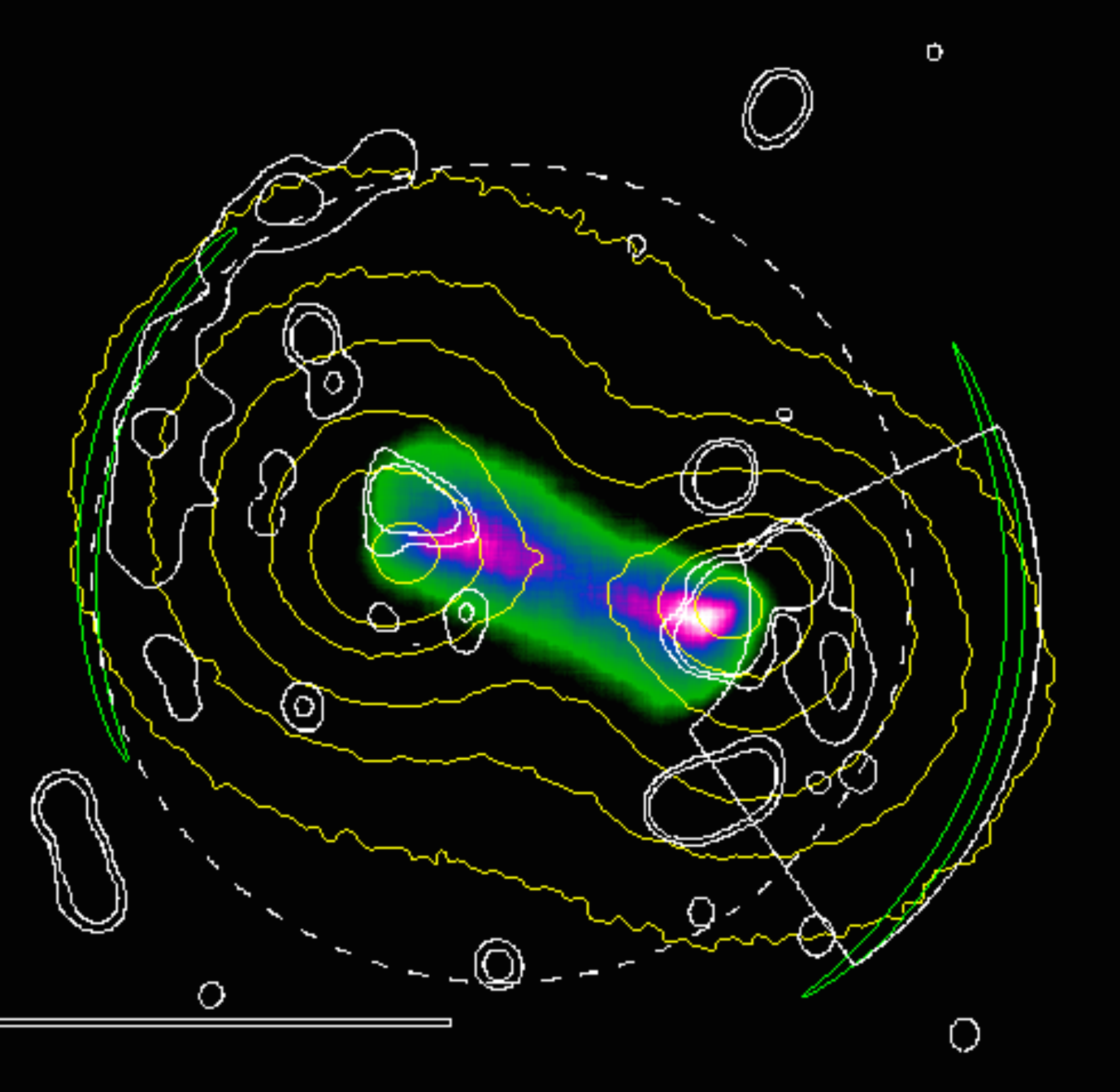} 
\includegraphics[width=.325\textwidth]{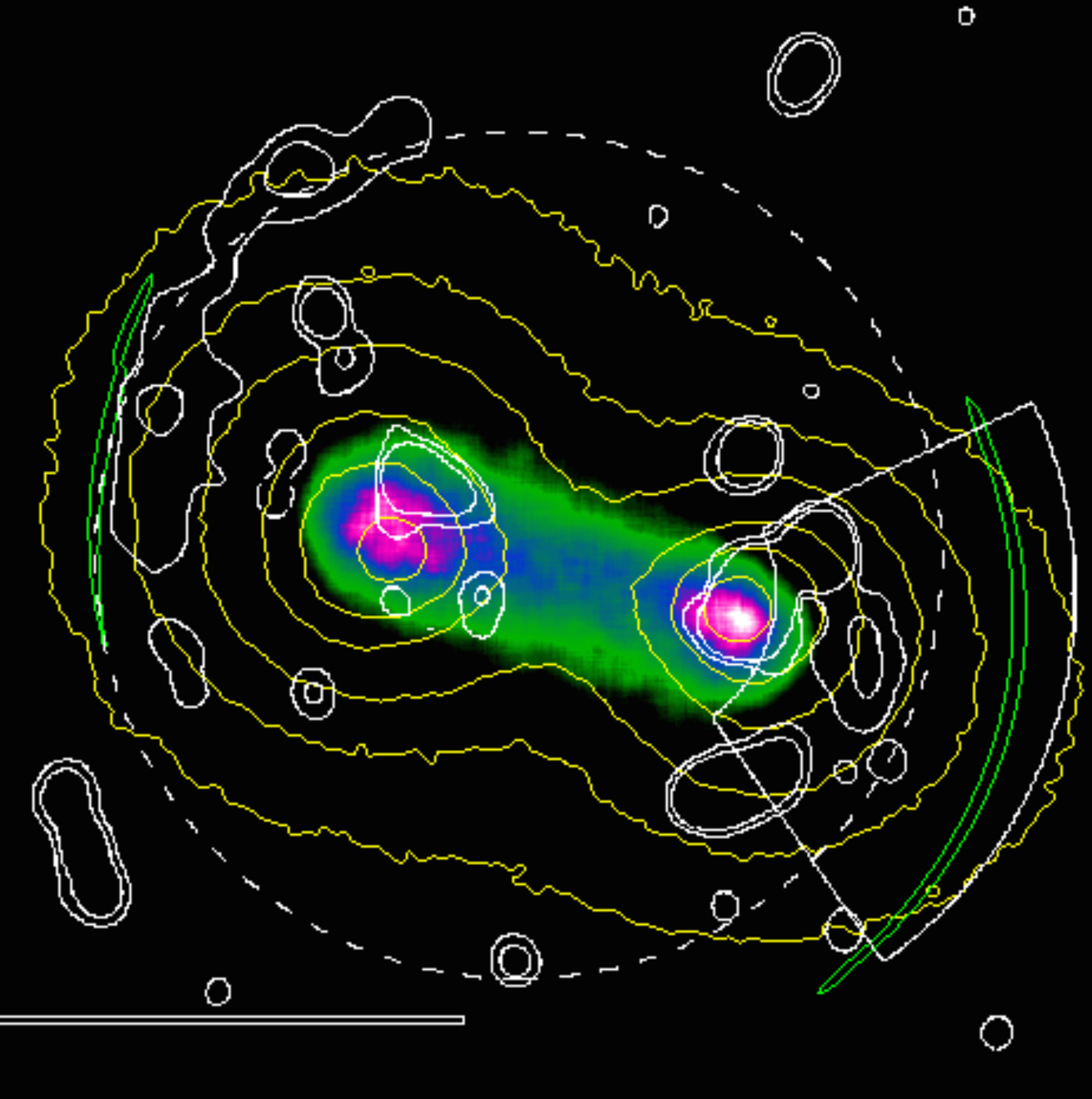} 
\includegraphics[width=.33\textwidth]{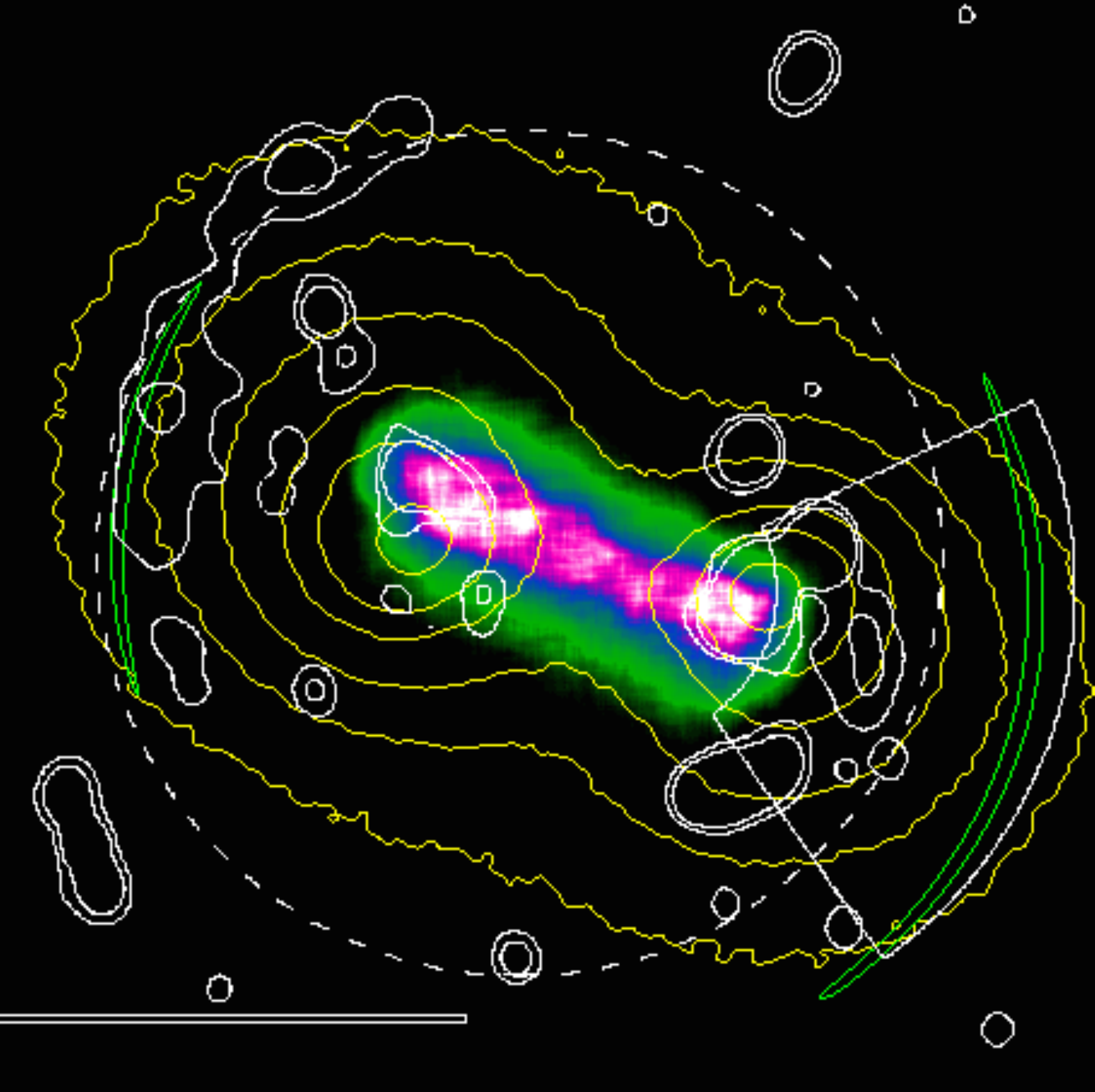}  
\includegraphics[width=.327\textwidth]{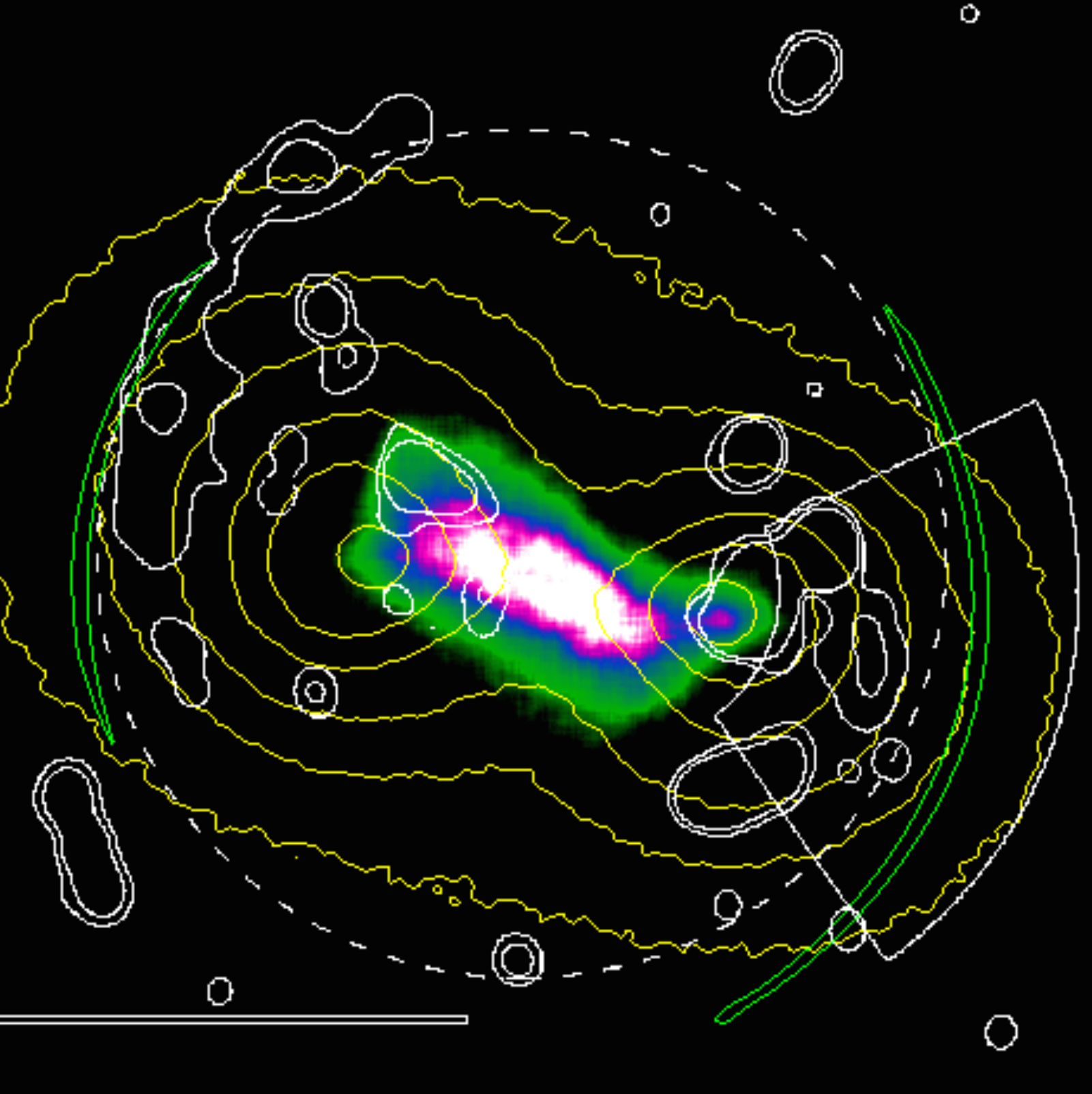} 
\includegraphics[width=.335\textwidth]{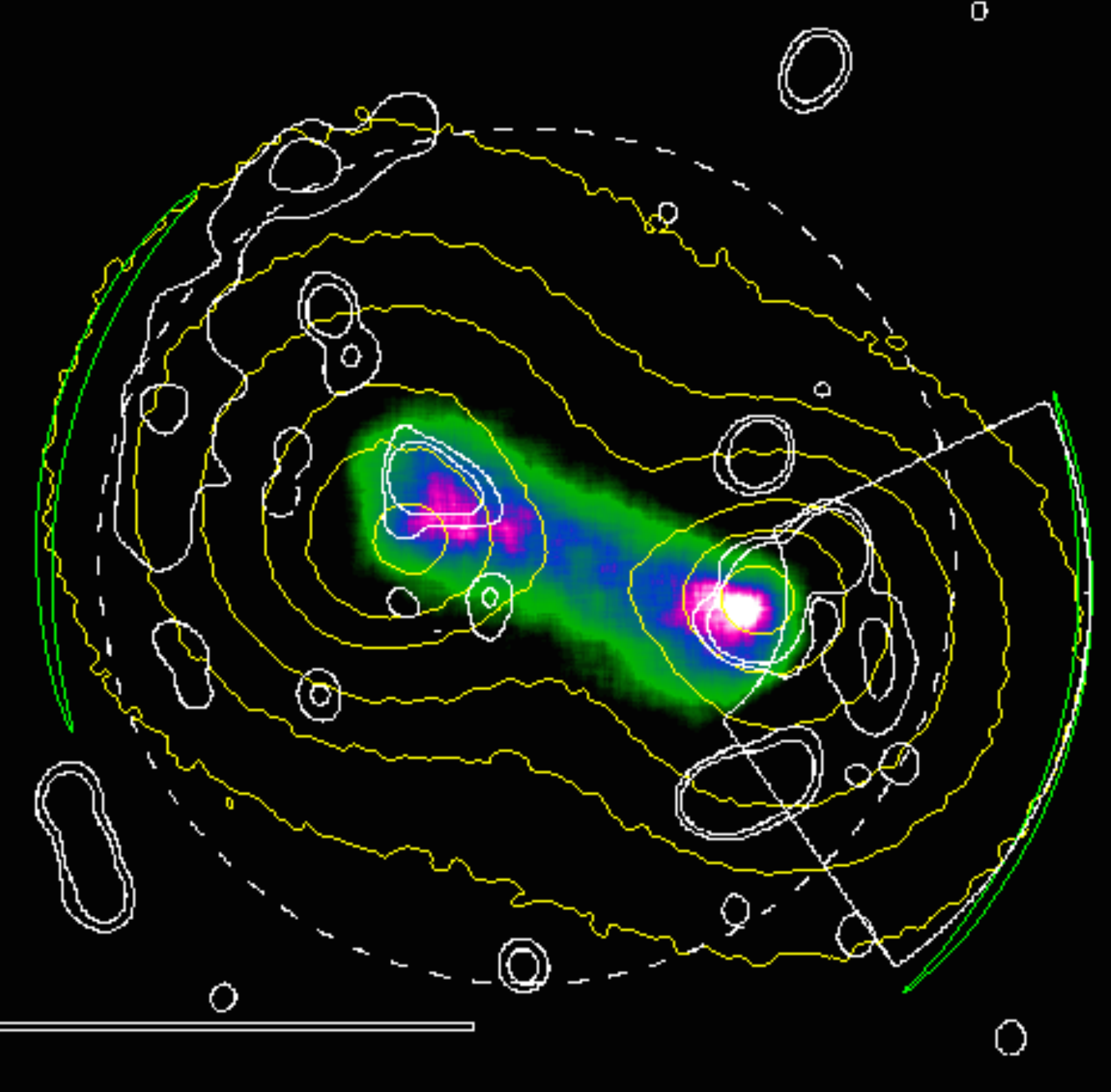} 
\includegraphics[width=.333\textwidth]{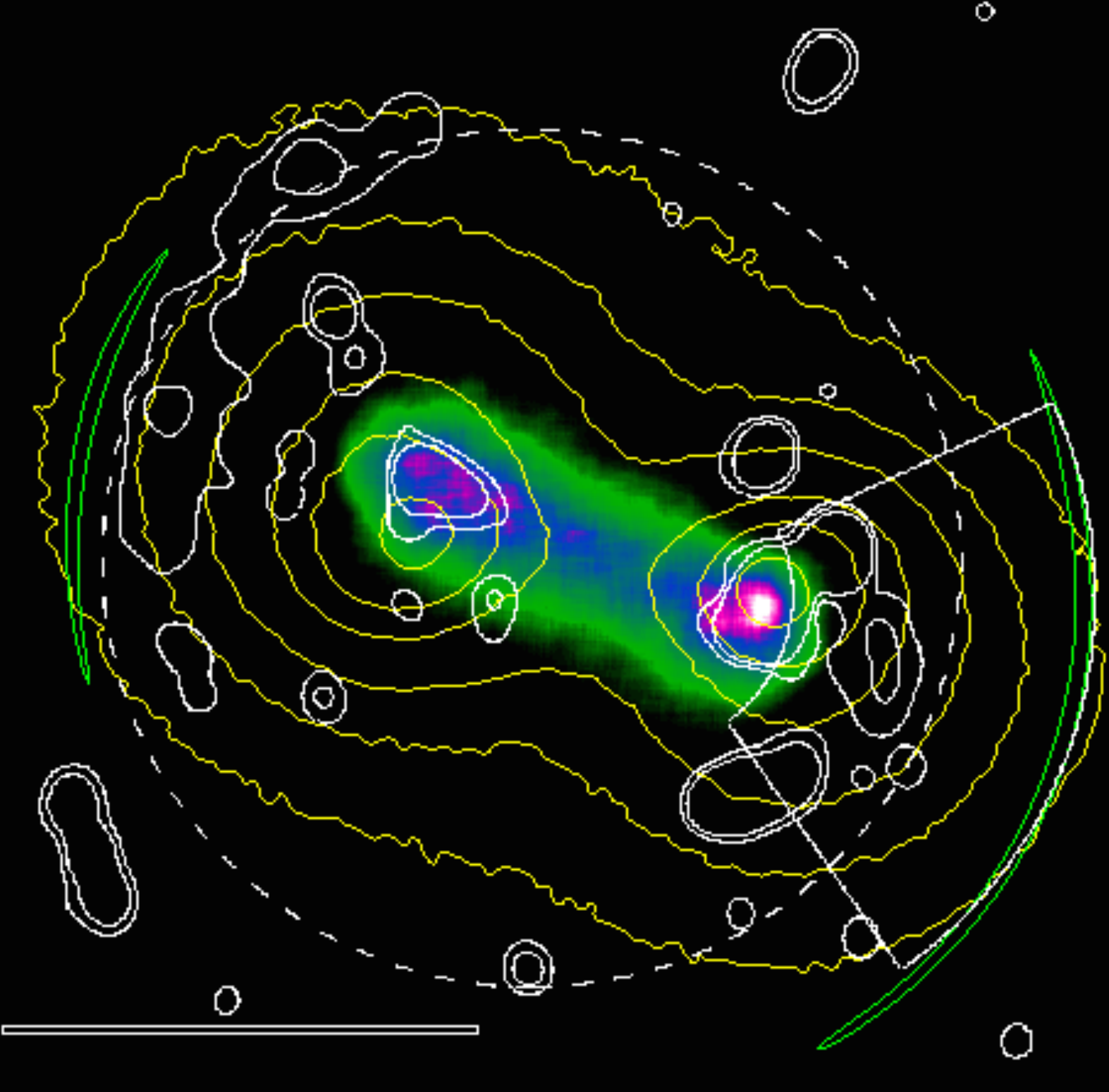}  
\caption{
Samples of simulations which do not fully resemble the X-ray morphology of \ZWCLEIGHT.
The color code is the same as in the second panel in Figure~\ref{F:OBSMOD}
Left to right: 
First row:
1st and 2nd panel show runs with  P = 300 and 500 kpc (P20V18 and P50V18)
Second row:
1st and 2nd panel show  V$_{in}$ = 1500 and 2000 \KMSEC;
The 3rd panel in the first and second row display the best model,
but before and after the best-fit epoch 
(412 Myr and 444 Myr after the first core passage; run P40V18).
The best-fit epoch is t$_{obs}$ = 428 Myr after the first core passage.
See Table~\ref{T:TABLE1} for a list of initial parameters for the runs.
\vspace{0.3 cm}
\label{F:SIMNOMATCH}
}
\end{figure*} 

\section{Results and discussion}
\label{S:RESULTS}

\subsection{Dynamical Model for \ZWCLEIGHT}
\label{SS:DEPROJECT}

The second panel in Figure~\ref{F:OBSMOD} shows a simulated X-ray color image 
of our best-fit model at the epoch of t$_{b} = 428$ Myr after the first core passage 
for \ZWCLEIGHT, run P400V18B with infall velocity, $V_{in} = 1800$ \KMSEC,
impact parameter, $P = 400$ kpc, and masses  
$M_{vir1} = 7$ and $M_{vir2} = 5 \times$\MSUNFOUR (main and infalling cluster).
The yellow contours represent the projected dark matter distribution from our simulation.
The white contours are based on radio observations, 
the white dashed circle and the horizontal bar represent R$_{500}$ 
and a physical length of 1 Mpc (from \citealt{GolovichET2017}; as in the firtst panel).
The viewing (rotation) angles were chosen the following way.
First we rotated the system with an angle, $\varphi = 30$\DEG (``roll angle''),
around the axis connecting the two dark matter centers (rotation around the $y$ axis)
then we rotated this axis with an angle, $\theta = 31$\DEG, out of the $x-y$ plane.
$\theta$ was chosen to provide a projected distance between the 
two dark matter centers, $D = 940$ kpc, to match the positions of the 
observed centers based on weak lensing mass reconstruction,
the roll angle was chosen to find the best match with the observed X-ray morphology 
\citep{GolovichET2017}.
We choose the output (epoch) which could be rotated in a way that the position of the 
back shock is near the eastern edge of the observed long radio relic in the east 
of the X-ray peak and the bow shock in the west is not inside the radio relic associated with it,
as radio observations suggest \citep{WeerenET2011AA528}.
The shocks in our simulated images were located based on projected pressure gradients.
A detailed description of our method to generate mock X-ray and mass surface density 
images can be found in \cite{MolnarBroadhurst2015}.

%
%
\begin{figure*}[t]
\includegraphics[width=.247\textwidth]{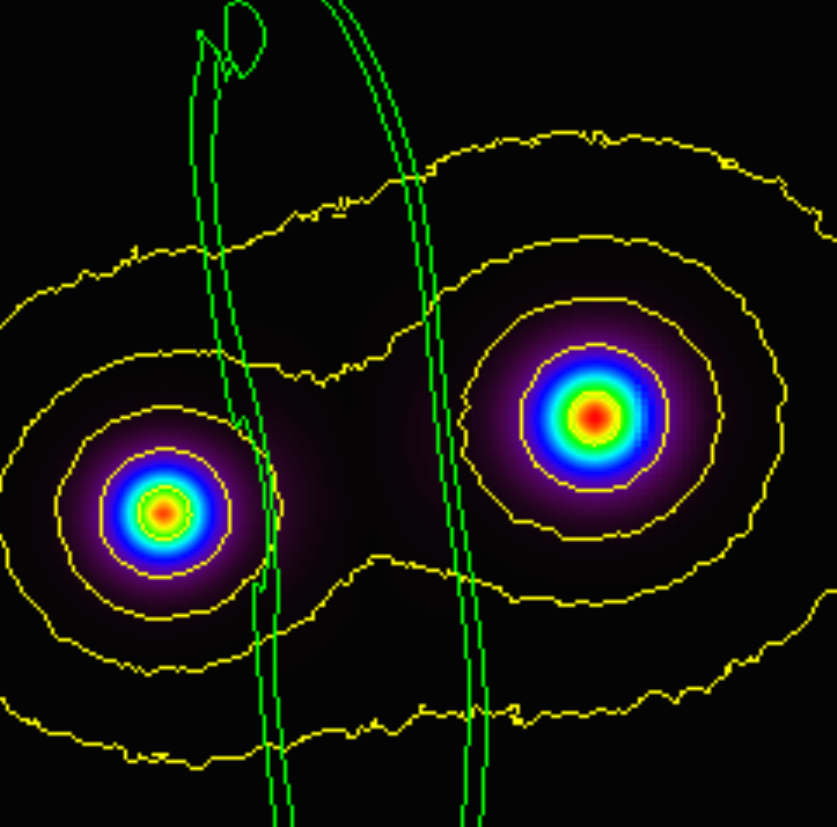}
\includegraphics[width=.247\textwidth]{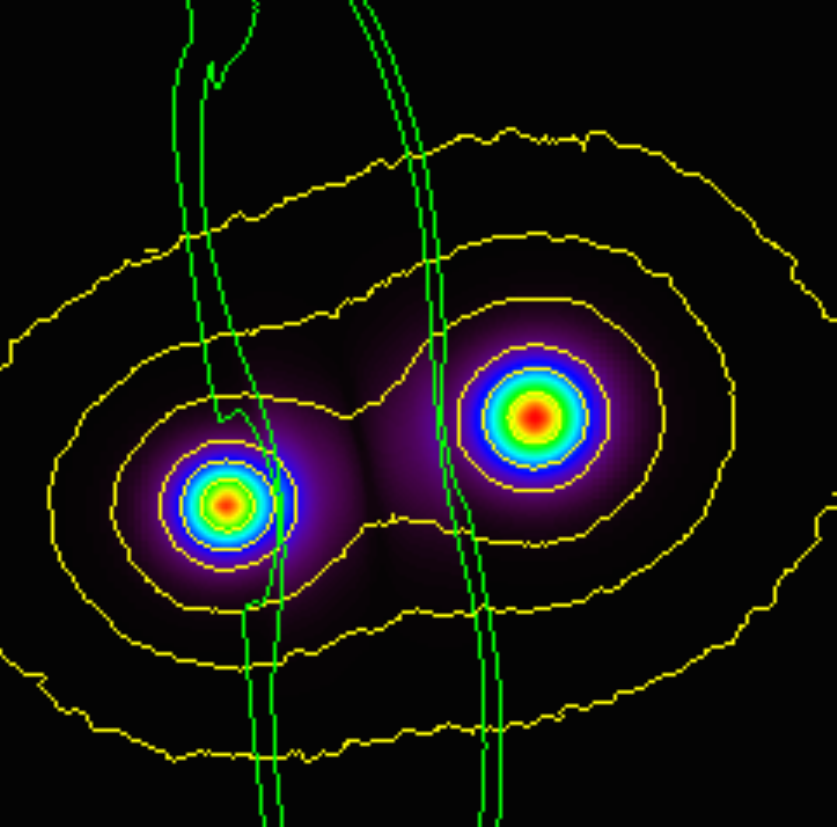}
\includegraphics[width=.247\textwidth]{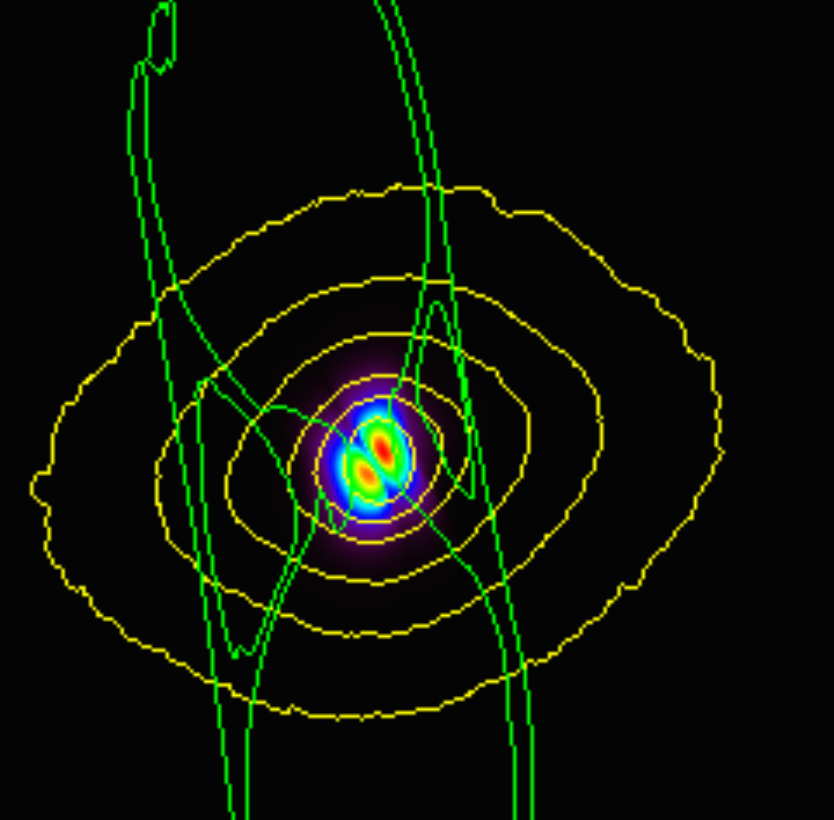}  
\includegraphics[width=.247\textwidth]{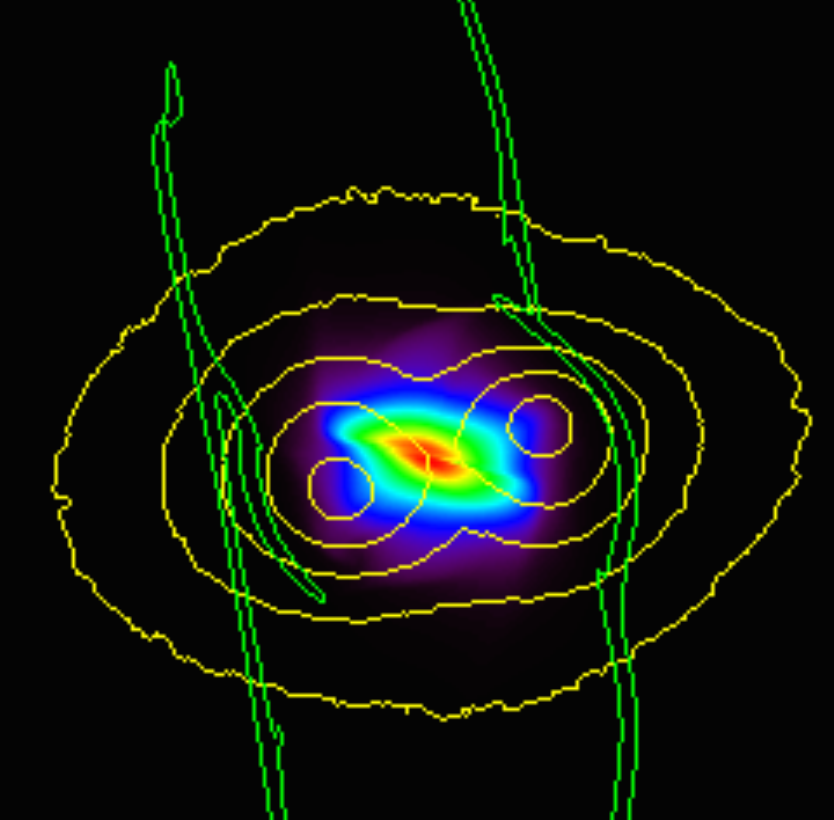}
\includegraphics[width=.247\textwidth]{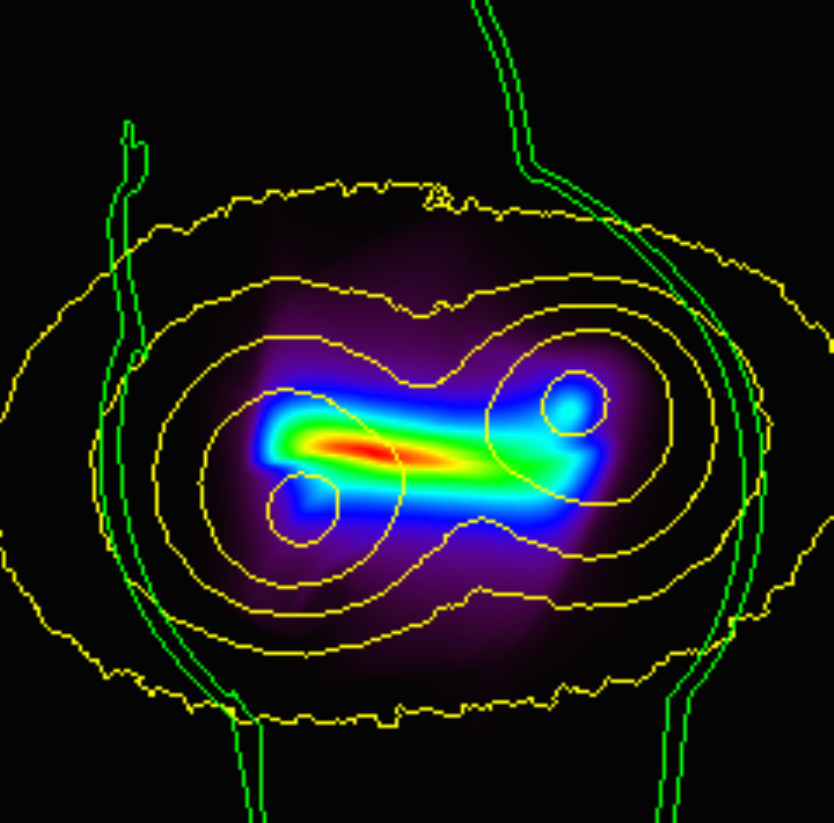}
\includegraphics[width=.247\textwidth]{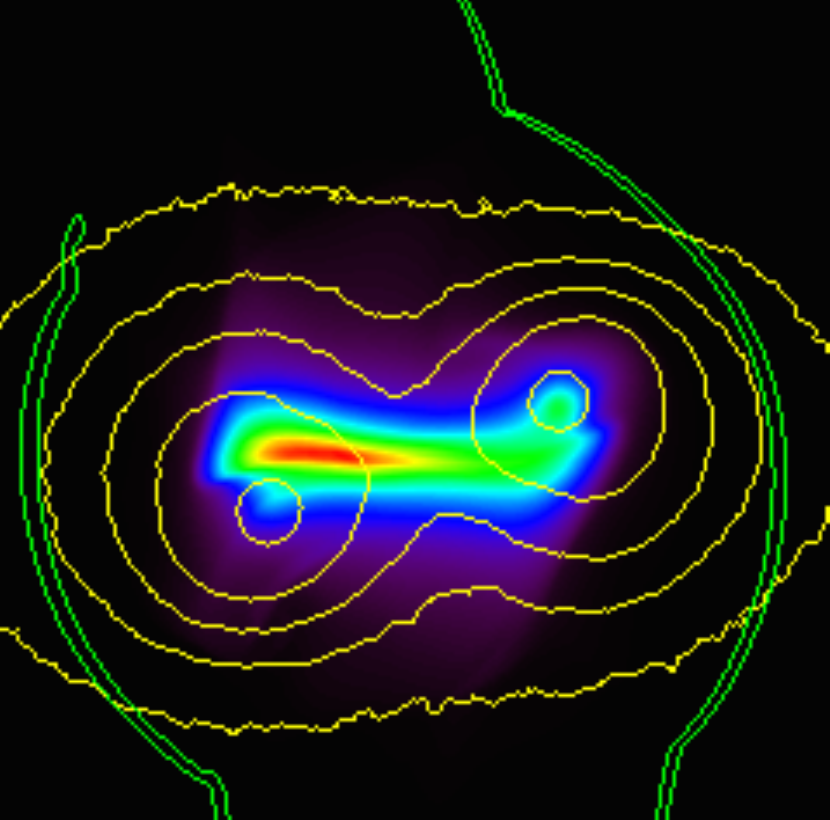}
\includegraphics[width=.247\textwidth]{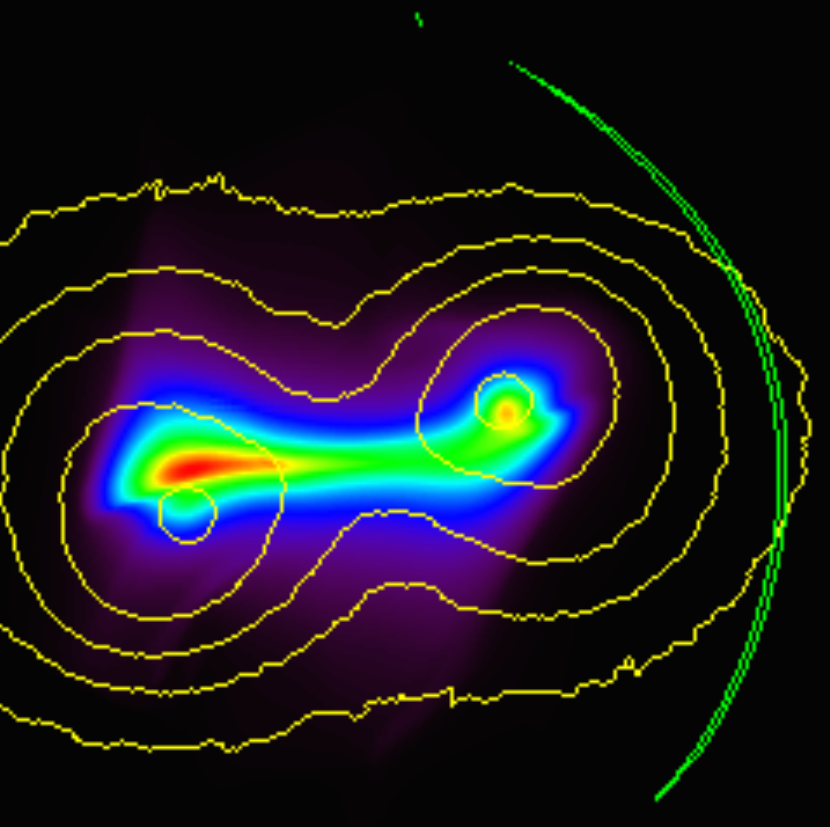}
\includegraphics[width=.247\textwidth]{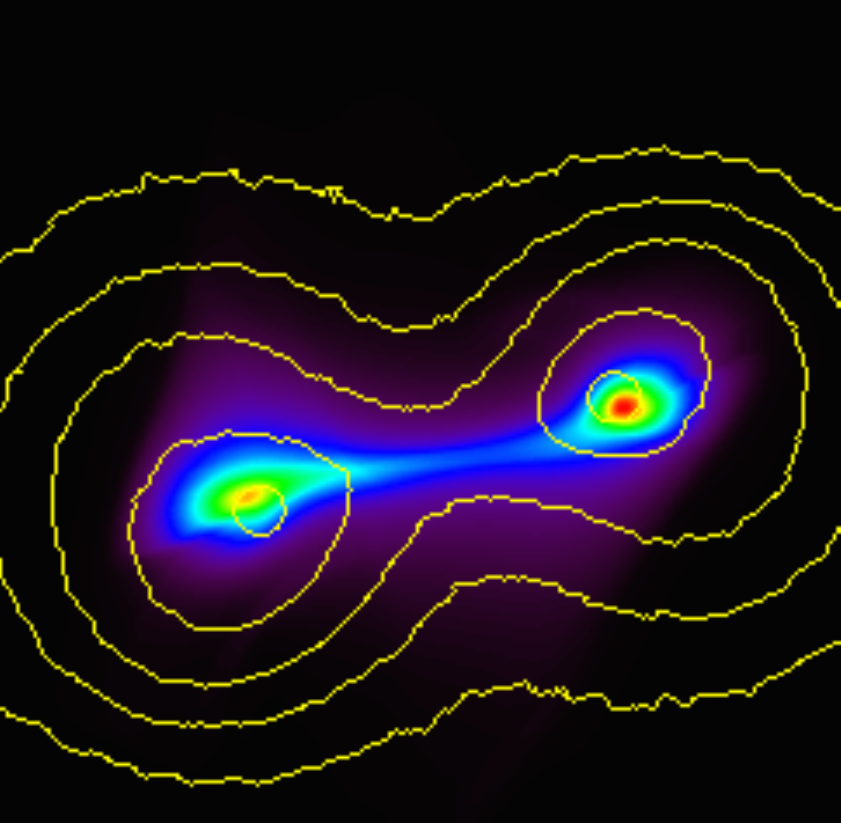}
\caption{
Merging shocks and the distribution of the mass 
surface density of dark matter (green and yellow contours)
overlaid on the X-ray emission (color image) as a function of time 
before and after the first core passage based on our best model for \ZWCLEIGHT
(run P400V18, see Table~\ref{T:TABLE1}).
We assumed that the collision occurs in the plane of the sky
(projections in the LOS). The panels are 3 Mpc on a side.
Left to right first and second row the epochs 
(relative to the first core passage; t$_0  = 0$)
are: $t$ = -475, -317, 0, 238, and $t$ = 396, 459, 555, 713 Myr.
The infalling cluster passes the main cluster from below moving east to west.
Panels in the first row: 
The first two panels in the first raw show epochs before the first core passage.
The 3rd panel represents the epoch of the first core passage.
The two dark matter peaks overlap at the center.
The 4th panel shows the phase right after the first core passage.
Panels in the second row show phases of the collision when there are
two X-ray peaks. 
The first two panels show two shocks ahead
of the cluster centers moving outwards.
The 3rd panel shows the phase when the back shock already ran out of
the gas of the infalling cluster. The bow shock is still moving outward
(towards west).
The 4th panel shows a later epoch, when both shocks ran out of the 
gas of the merging system, and the gas is falling back to the cores of the 
dark matters of the two components.
\vspace{0.5 cm}
\label{F:OUTSHOCKS}
}
\end{figure*} 
\vspace{1.0 cm}

In Figure~\ref{F:SIMNOMATCH}, we show images of models for 
\ZWCLEIGHT, which do not satisfy our requirements for a good 
match with the data.
In this figure the color images represent the simulated X-ray 
surface brightness, the color code for the contours is the same as
in the second panel in Figure~\ref{F:OBSMOD}.
The procedure was the following. First we aligned the position of the X-ray 
peak associated with the infalling cluster with that of the observed, then
we choose the projection angles to match 
the distances and position angles between the dark matter centers to match 
those derived from weak lensing observations.
In the third step, if it was possible possible, we selected cases where the position 
of the back shock is near the eastern edge of the long radio relic in the east of the X-ray peak.

The first and second panels in the first row show runs with only 
the impact parameters changed to P = 300 and 500 kpc (runs P20V18 and P50V18)
from the best model (run P400V18 with P = 400 kpc).
The X-ray morphology of P20V18 (first panel) is too elongated 
along the line connecting the two dark matter centers.
Run P50V18 displays two X-ray peaks with large separation and
very small offset from the dark matter centers.
The first and second panel in the second row 
show simulations changing only the relative (infall) velocity of the infalling cluster 
to V$_{in}$ = 1500 and 2000 \KMSEC (runs P40V15 and P40V20)
to bracket our best model whith V$_{in}$ = 1800 \KMSEC (run P400V18).
In the first panel, the very bright X-ray peak is associated to the main cluster,
not to the infalling cluster as observed, in the second panel the back shock
is farther than observed.

The third panel in the first and second row display the best model 
but at $t$ = 412 Myr and 444 Myr after the first core passage
(runs P40V18T1 and P40V18T2), before and after the best-fit epoch.
The best-fit epoch is t$_{obs}$ = 428 Myr after the first core passage (run P40V18).
See Table~\ref{T:TABLE1} for a list of initial parameters for the runs.
The third panel in the first row shows two bright X-ray peaks with enhanced
emission from the tidal bridge between them, which differs from the data where the
X-ray peaks have a large brightness ratio and a less enhanced bridge between them.
The third panel in the second row shows two X-ray peaks with
a large separation and a small offset for both X-ray peaks,
again differing significantly with respect to the observed morphology.

We conclude from of our suite of $N$-body/\-hydro\-dynamical simulations 
that \ZWCLEIGHT is viewed at about 428 Myr after first core passage,
and that the infalling cluster has a mass M$_{vir;2} = 5\pm 0.5\,$\TMSUNFOUR 
moving to the west, disrupting the gas of the main cluster with mass
M$_{vir;1} = 7\pm 0.5\,$\TMSUNFOUR, so that it lies to the west of the 
the main cluster at the observed epoch. Our model also clearly confirms that 
the disrupted gas of the main cluster is offset from its the dark matter center
as the data seem to indicate.   
Our simulations clearly demonstrate that the merging cluster, \ZWCLEIGHT, 
is in the outgoing phase just after the first core passage, before the first
turnover. The gas and dark matter associated with the two components 
are moving outward (the infalling cluster moving to the east, the 
main cluster to the west; see fist and second panels in Figure~\ref{F:OBSMOD}).

These results are in broad qualitative agreement with the results of 
\cite{GolovichET2017}, but clearly prefer a recent post collision epoch and 
exclude their later, post collision epoch option of 1300 Myr, as by then the shock 
fronts predicted by our model will have long left the system.
Also, our best model has significantly larger mass, $1.2 \times 10^{15} \,M_\odot$,
and a smaller mass ratio, 1.4, than those suggested by Golovich et al.
($6.9 \times 10^{14} \,M_\odot$ and 4.75).

We performed a simulation with initial conditions suggested by \cite{GolovichET2017}
(run P10V15G). We show the result in the third panel in Figure~\ref{F:OBSMOD}.
There is only one X-ray peak, which is associated with the gas of the main cluster.
We chose a projection, which provides a roughly match to the positions of the two 
merger shocks with those observed. However, there is no X-ray peak associated 
with the infalling cluster, contrary to the observations. 
The gas of the infalling cluster has been stripped off as a result of the relatively large  
velocity of the infalling cluster (1500 \KMSEC) and the small impact parameter (100 kpc).
Note, that \cite{GolovichET2017} assumed zero impact parameter, which would
make it even easier to strip all the gas from the infalling cluster. We chose a finite,
but small impact parameter since the observed X-ray morphology is not symmetric.

\subsection{Outgoing Merging Shocks in \ZWCLEIGHT}
\label{SS:SHOCKPROPERTIES}

The properties of merging shocks were studied in detail by making use of  
$N$-body/\-hydro\-dynamical simulations 
\citep[e.g.,][]{MolnarBroadhurst2017,HaET2017arXiv170605509}.
In this section we study the evolution of merging shocks as a function of time
around the first core passage before the first turnover using our best
solution for \ZWCLEIGHT as an example.

The evolution of merging shocks is illustrated in Figure~\ref{F:OUTSHOCKS}.
The panels in this figure show a color image of the X-ray emission
with contours of the dark matter distribution (yellow) and the
merging shocks (green) superimposed. 
We used a projection along the $z$ axis, i.e., we assumed that $z$ coincides with
the line of sight (LOS), and the collision takes place in the plane of the sky ($x,y$ plane).
The infalling cluster passes the main cluster center from below moving east to west
(second panel in the first row; run P40V18).
The time is measured in Myr relative to the first core passage (t$_0  = 0$ Myr; 
left to right first: $t$ = -475, -317, 0, 238, Myr, and 
second row: $t$ = 396, 459, 555, 713 Myr).

The first two panels in the first row show epochs before the first
core passage, when only the outer regions of the gas in the two clusters collide.
The shocks move ahead of the two clusters generating them
in the same direction as their respective dark matter
(the shock on the east moves to west, the shock on the west moves to east). 
The third panel represents the epoch of the first core passage.
The two dark matter peaks overlap at the center; they are slightly 
ahead of the X-ray peaks. The structure of the merging shocks change,
multiple shocks are generated due to the collision of the more dense
gas in the merging clusters.
The fourth panel shows a phase right after the first core
passage, when there is only one X-ray peak and the bow shock on the
west in the gas of the main cluster is moving to the west ahead of the infalling 
cluster. The back shock in the east propagates to the east in the gas of the 
infalling cluster.
Panels in the second row show phases after the first core passage, 
when there are two X-ray peaks. 
The first two panels show two shocks ahead 
of the cluster centers moving outward.
The third panel shows a phase when the back shock already ran out of
the gas of the infalling cluster. The bow shock is still moving outward
(towards west).
The fourth panel shows a later epoch, when both shocks ran out of the 
gas of the merging system, and the gas is falling back to the cores of the 
dark matter of the two components.

We study these outgoing merging shocks in detail because of
their importance in detrerming the dynamical state of the
merging cluster, and the phase of the collision.
The phase of the collision is important for particle acceleration models, 
because the physical properties of the shock depend on it, and these properties 
have a large impact on particle acceleration
(e.g., \citealt{FujitaET2016,KangRyu2016,StroeET2014MNRAS445}),
as well as testing cosmological models \citep[e.g.][]{Molnar2015}.

The reason why the outgoing shocks constrain the phase of the 
collision well right after the first core passage in the outgoing phase
is that they move fast and run out of the gas as they propagate 
with a high speed in the low density gas at the outskirts of the system. 
Because of these relatively high outer shock speeds, these shocks 
can (re)accelerate particles and generate luminous radio relics for a limited time.
As a consequence shocks and bright relics can only be expected to 
be detected in a pair of merging clusters relatively soon after the first core passage,
before the first turnaround, after which the relics become fainter without a shock 
to provide additional energy.

In general, as we see here, if the observed shocks were to lie in between the two 
X-ray peaks associated with relaxed cluster centers, 
then we would conclude that the merging system is being viewed before 
the first core passage, as shown in the first panel in Figure~\ref{F:OUTSHOCKS}, 
such as the case of, e.g., Abell 1750 \citep{Molnar2013ApJ779}.
Disturbed X-ray morphology displaying with one or two X-ray peaks within one or two
shocks instead suggests a system caught after first core passage 
(e.g., \citealt{MolnarBroadhurst2015,MastBurk08MNRAS389p967}.

Figure~\ref{F:OUTSHOCKS} also illustrates how fast the outgoing 
shocks propagate in a merging cluster similar to \ZWCLEIGHT. 
After $\sim$500 Myr, both the bow shock and the back shock
propagating in the gas of the main and the infalling cluster 
ran out of the system (panel 4 in Figure~\ref{F:OUTSHOCKS}.
The velocity relative to the ambient gas of the bow shock 
(the shock on the west moving to west) is 4500 \KMSEC, 
the back shock propagates faster (to the east) with 4900 \KMSEC. 
At these speeds, the merging shocks (bow and back shocks)
would run out of the gas in 470 and 380 Myr after the first core passage, 
but our numerical simulations suggest that they run out in 618 and 522 Myr.
The turnover occurs 1.5 Gyr after the first core passage,
thus, well before the turn over, both merger shocks leave the system.
This is a general feature of merging shocks more colliding clusters
with moderate mass ratios and infall velocities $\simgreat 1000$ \KMSEC.

We derive the Mach numbers for the merging shocks based on our best model
using the temperature jump at the shocks, as it is done using temperature
based on X-ray observations.
The Rankine-Hugoniot jump conditions provide the connection between 
the temperature jump, $T_2/T_1$, and the Mach number, 
\begin{equation}
   \frac{T_2}{T_1} = \frac{5 \MACH^2 + 14 \MACH^2 -3}{16 \MACH^2}
,
\label{E:MACHT2T1}
\end{equation}
where $T_1$ and $T_2$ are the pre- and post-shock temperatures
(e.g., \citealt{MolnarBroadhurst2017,AkamatsuET2015}; 
for a review see \citealt{MarkevitchVikhlinin2007}).
Using the temperature jump from our simulations at the shocks 
in Equation~\ref{E:MACHT2T1}, 
we obtain Mach numbers for the merging shocks
directly from the physical (not the observed) temperature ratios
(the bow shock and back shock to the west and east):
$\MACH_{w,simu} = 5.5$ and $\MACH_{e,simu} = 6.5$.

\subsection{Comparing \ZWCLEIGHT to the Bullet Cluster}
\label{SS:COMPARISON}

It has been suggested that \ZWCLEIGHT is an older, less massive 
version of the Bullet Cluster \citep{GolovichET2016ApJ831}, 
where the less massive infalling cluster has pushed the gas of the main 
cluster out of equilibrium and, as a result, the X-ray morphology, 
which traces the gas of the main cluster,
is irregular (X-ray feature in the east in the first panel in 
Figure~\ref{F:OBSMOD}).
However, our hydrodynamical simulations suggest that the 
cometary-like X-ray peak in \ZWCLEIGHT near the center of mass
of the smaller western cluster (see first panel in Figure~\ref{F:OBSMOD})
marks the gas density maximum of the infalling cluster,
similar to El Gordo \citep{MolnarBroadhurst2015}, 
unlike the bullet cluster, in which the wedge-shaped bright X-ray feature
marks a contact discontinuity.
The shock in the Bullet cluster is a faint X-ray feature 
ahead of the merging shock to the west 
(e.g., \citealt{MarkevitchVikhlinin2007}).
Based on our best model, \ZWCLEIGHT is only 428 Myr after the first
core passage, and in a much earlier phase than the turnover, 1.5 Gyr,
like the Bullet cluster.

Our results suggest instead that, because of a larger impact parameter
for the collision in \ZWCLEIGHT and the lower infall velocity relative to the
Bullet cluster, the ram pressure cannot fully hold back
the gas relative to the dark matter in the infalling cluster, 
as opposed to the case of the Bullet cluster.
Instead, the infalling cluster in \ZWCLEIGHT passes relatively unhindered 
by the gas of the main cluster since it does not penetrate through the 
central dense gas of the main cluster.

\section{Conclusions}
\label{S:CONCLUSIONS}

We have performed a set of self-consistent $N$-body/\-hydro\-dynamical simulations 
based on \FLASH, a publicly available AMR code, augmenting our existing library 
of binary cluster encounters generated in our previous work, in order to study the 
particular dynamics during the collision of the merging cluster \ZWCLEIGHT.
We have modeled \ZWCLEIGHT as a binary merger, constraining the initial
parameters using gravitational lensing, X-ray morphology and observations
of radio relics.
Unfortunately the merger shock positions cannot be determined
from the \CHANDRA observations without longer X-ray exposures.
Therefore, we refer to the positions of the pair of opposing distinctively 
polarized radio relics that we can assume to have resulted from the 
two predicted outgoing shocks and a simple geometrical argument 
(see Section~\ref{SS:DEPROJECT}) to limit the current shock locations. 
The detailed X-ray morphology and the locations 
of the two lensing centroids were used to constrain the impact parameter and 
the infall velocity of the collision, as well as the viewing angle.

We have demonstrated, that the outgoing shocks travel fast 
(4000-5000 \KMSEC) in the low density outer gas of the two subclusters, 
and, therefore, their positions relative to those of the mass peaks 
can be used to accurately derive the phase of the collision.
Thus, merging shocks can (re)accelerate particles and generate relics
only for a limited time $\simless 5\times 10^8$ yr, so that shocks and bright relics can be 
detected in a merging system soon after the first core passage, before the first turnaround. 
After that time the relics become fainter after no shocks feed them as the 
electrons loose energy. 
This point has been unappreciated in earlier work \citep{GolovichET2017,NgET2015}
where later stage merging has been entertained without the guidance of 
hydrodynamical simulations like those presented in this paper.

Based on our $N$-body/\-hydro\-dynamical simulations, 
we derive an impact parameter of $400 \pm 100$ kpc, 
and an infall velocity of $1800 \pm 300\,$\KMSEC, 
with virial masses of M$_{vir;1} = 7\pm 0.5\,$\TMSUNFOUR and 
M$_{vir;2} = 5\pm 0.5\,$\TMSUNFOUR for the main and infalling cluster respectively.
We find that \ZWCLEIGHT is observed about 430 Myr after the first core passage.
Our simulations clearly demonstrate that \ZWCLEIGHT
is currently in the outgoing phase, well before the first turnover,
otherwise the forward and reverse shock fronts would have long run out
of the system to the east and west.

Our numerical simulations represent the first attempt to model the 
newly discovered Bullet-cluster-like merging system \ZWCLEIGHT 
using self-consistent $N$-body/\-hydro\-dynamical simulations. 
Previously \ZWCLEIGHT was modeled by 
\cite{GolovichET2017} using their method based on a model assuming 
fixed NFW profiles for the dark matter distribution for the merging subclusters 
assuming zero impact parameter and ignoring the gas components \citep{Dawson2013ApJ}.
However, their model cannot distinguish between phases of 
outgoing or infalling after the first turnover.
Our full self-consistent simulation containing dark matter and gas can constrain 
the impact parameter, the phase of the collision,
and the viewing angle with the location of the merging shocks
to provide a both check on the apparent interpretation of this system
as a binary encounter and to provide reliable estimates of the 
basic masses, velocities and the age and orientation of the system.

The degree of agreement we find between all the reliable observables of the binary 
merging cluster \ZWCLEIGHT and those based on our best model derived from 
$N$-body/\-hydro\-dynamical simulations along with self-consistent simulations of 
other merging clusters provides further strong evidence that dark matter is effectively
collisionless on large scales.
These results support the remarkable insight into this question initially gained by the Bullet cluster.
This self consistency calls into question other claims and theories that advocate 
modified gravity, where the aim is to ``emulate'' dark matter,
simply, because the lensing contours indicating the location of the gravitational
potential do not follow the dominant baryonic material that is composed of gas. 
Instead the detailed gas distribution relative to the lensing data indicates the contrary, 
that dark matter dominates and it is collisionless to within the precision of the data.

\acknowledgements
The code \FLASH\ used in this work was in part developed by the
DOE-supported ASC/Alliance Center for Astrophysical Thermonuclear
Flashes at the University of Chicago.  
We thank the Theoretical Institute for Advanced Research in Astrophysics,
Academia Sinica, for allowing us to use its high performance computer facility 
to carry out our simulations.


%
%
\bibliographystyle{apj}

\begin{thebibliography}{99}




\bibitem[Agertz et al.(2007)]{AgertzET2007}
 Agertz, O., Moore, B., Stadel, J., et al.\ 2007, \mnras, 380, 963 



\bibitem[Akamatsu et al.(2015)]{AkamatsuET2015}
 Akamatsu, H., van Weeren, R.~J., Ogrean, G.~A., et al.\ 2015, \aap, 582, A87 



 
\bibitem[{{Barrena} {et~al.}(2009){Barrena}, {Girardi}, {Boschin}, \& {Das{\'{\i}}}}]{BarrenaET2009}
 {Barrena}, R., {Girardi}, M., {Boschin}, W., \& {Das{\'{\i}}}, M. 2009, \aap, 503, 357


\bibitem[Basu et al.(2016)]{BasuET2016AA591}
 Basu, K., Vazza, F., Erler, J., \& Sommer, M.\ 2016, \aap, 591, A142 

\bibitem[Botteon et al.(2016)]{BotteonET2016}
 Botteon, A., Gastaldello, F., Brunetti, G., \& Kale, R.\ 2016, \mnras, 463, 1534 





\bibitem[{{Brada{\v c}} {et~al.}(2008){Brada{\v c}}, {Allen}, {Treu},{Ebeling}, {Massey}, {Morris}, 
                                                                                                {von der Linden}, \& {Applegate}}]{BradacET2008}
 {Brada{\v c}}, M., {Allen}, S.~W., {Treu}, T., {Ebeling}, H., {Massey}, R., {Morris}, R.~G., 
 {von der Linden}, A., \& {Applegate}, D. 2008, \apj, 687, 959



 
 \bibitem[Bouillot et al.(2015)]{BouillotET2015}
  Bouillot, V.~R., Alimi, J.-M., Corasaniti, P.-S., \& Rasera, Y.\ 2015, \mnras, 450, 145 



\bibitem[Cai et al.(2009)]{CaiET2009}
 Cai, Y.-C., Angulo, R.~E., Baugh, C.~M., et al.\ 2009, \mnras, 395, 1185 




\bibitem[Clowe \etal(2006)]{CloweET2006}
 Clowe, D., Brada{\v c}, M., Gonzalez, A.~H., \etal, 2006, \apjl, 648, L109 
 
 


\bibitem[{{Clowe} {et~al.}(2004){Clowe}, {Gonzalez}, \& {Markevitch}}]{CloweET2004}
{Clowe}, D., {Gonzalez}, A., \& {Markevitch}, M. 2004, \apj, 604, 596


\bibitem[Dawson(2013)]{Dawson2013ApJ}
 Dawson, W.~A.\ 2013, \apj, 772, 131 

\bibitem[{{Dawson} {et~al.}(2012){Dawson}, {Wittman}, {Jee}, {Gee}, {Hughes}, {Tyson}, {Schmidt}, 
                           {Thorman}, {Brada{\v c}}, {Miyazaki}, {Lemaux}, {Utsumi}, \& {Margoniner}}]{DawsonET2012}
 {Dawson}, W.~A., {et~al.} 2012, \apjl, 747, L42



\bibitem[Donnert et al.(2017)]{Donnert2017MNRAS471}
 Donnert, J.~M.~F., Beck, A.~M., Dolag, K., \& R{\"o}ttgering, H.~J.~A.\ 2017, \mnras, 471, 4587 





\bibitem[En{\ss}lin(1999)]{Ensslin1999}
 En{\ss}lin, T.~A.\ 1999, Diffuse Thermal and Relativistic Plasma in Galaxy Clusters, 275 
  

\bibitem[Feretti \etal(2012)]{FerettiET2012}
 Feretti, L., Giovannini, G., Govoni, F., \& Murgia, M., 2012, \aapr, 20, 54 



\bibitem[Fryxell et al.(2000)]{Fryxell2000ApJS131p273}
 Fryxell, B., et al.\ 2000, \apjs, 131, 273 




\bibitem[Fujita et al.(2016)]{FujitaET2016}
 Fujita, Y., Akamatsu, H., \& Kimura, S.~S.\ 2016, \pasj, 68, 34 





\bibitem[Golovich et al.(2016)]{GolovichET2016ApJ831} 
Golovich, N., Dawson, W.~A., Wittman, D., et al.\ 2016, \apj, 831, 110 


\bibitem[Golovich et al.(2017)]{GolovichET2017}
 Golovich, N., van Weeren, R.~J., Dawson, W.~A., Jee, M.~J., \& Wittman, D.\ 2017, \apj, 838, 110 


\bibitem[Ha et al.(2017)]{HaET2017arXiv170605509}
 Ha, J.-H., Ryu, D., \& Kang, H.\ 2017, arXiv:1706.05509 



\bibitem[Hoang et al.(2017)]{HoangET2017}
 Hoang, D.~N., Shimwell, T.~W., Stroe, A., et al.\ 2017, \mnras, 471, 1107 


\bibitem[Kang \& Ryu(2016)]{KangRyu2016}
 Kang, H., \& Ryu, D.\ 2016, \apj, 823, 13 





\bibitem[Kierdorf et al.(2017)]{KierdorfET2017AA}
 Kierdorf, M., Beck, R., Hoeft, M., et al.\ 2017, \aap, 600, A18 


\bibitem[Kraljic \& Sarkar(2015)]{KraljicSarkar2015}
 Kraljic, D., \& Sarkar, S.\ 2015, JCAP, 4, 050 



\bibitem[Lage \& Farrar(2014)]{LageFarrar2014ApJ787}
 Lage, C., \& Farrar, G.\ 2014, \apj, 787, 144



\bibitem[Lage \& Farrar(2015)]{LageFarrar2015JCAP}
 Lage, C., \& Farrar, G.~R., 2015, JCAP, 2, 038 

\bibitem[Lam et al.(2014)]{LamET2014ApJ797}
 Lam, D., Broadhurst, T., Diego, J.~M., et al.\ 2014, \apj, 797, 98 


\bibitem[Lee \& Komatsu(2010)]{LeeKomatsu2010ApJ718}
Lee, J., Komatsu, E.,  2010, \apj, 718, 60



\bibitem[{\L}okas \& Mamon(2001)]{LokasMamon2001MNRAS321}
 {\L}okas, E.~L., \& Mamon, G.~A.\ 2001, \mnras, 321, 155 

\bibitem[Ma, Ebeling \& Barrett(2009)]{MaET2009}
 Ma, C.-J., Ebeling, H., \& Barrett, E., 2009, \apjl, 693, L56 


\bibitem[Machado et al.(2015)]{MachadoET2015}
 Machado, R.~E.~G., Monteiro-Oliveira, R., Lima Neto, G.~B., \& Cypriano, E.~S.\ 2015, \mnras, 451, 3309 





\bibitem[{{Mahdavi} {et~al.}(2007){Mahdavi}, {Hoekstra}, {Babul}, {Balam}, \& {Capak}}]{MahdaviET2007}
 {Mahdavi}, A., {Hoekstra}, H., {Babul}, A., {Balam}, D.~D., \& {Capak}, P.~L. 2007, \apj, 668, 806
BarrenaET2009



\bibitem[Markevitch et al.(2004)]{MarkevitchET2004ApJ606}
 Markevitch, M., Gonzalez, A.~H., Clowe, D., et al.\ 2004, \apj, 606, 819 


\bibitem[Markevitch et al.(2002)]{MarkevitchET02}
 Markevitch, M., Gonzalez, A. H., David, L., et al. 2002, \apj, 567, L27



\bibitem[Markevitch \& Vikhlinin(2007)]{MarkevitchVikhlinin2007}
 Markevitch, M., \& Vikhlinin, A., 2007, \physrep, 443, 1 



\bibitem[Massardi et al.(2010)]{MassardiET2010}
 Massardi, M., Ekers, R.~D., Ellis, S.~C., \& Maughan, B.\ 2010, \apjl, 718, L23 



\bibitem[Mastropietro \& Burkert(2008)]{MastBurk08MNRAS389p967}
 Mastropietro, C., \& Burkert, A.\ 2008, \mnras, 389, 967 


\bibitem[Medezinski et al.(2016)]{MedezinskiET2016ApJ817}
 Medezinski, E., Umetsu, K., Okabe, N., et al.\ 2016, \apj, 817, 24 


\bibitem[{{Menanteau} {et~al.}(2012){Menanteau}, {Hughes}, {Sif{\'o}n},
  {Hilton}, {Gonz{\'a}lez}, {Infante}, {Barrientos}, {Baker}, {Bond}, {Das},
  {Devlin}, {Dunkley}, {Hajian}, {Hincks}, {Kosowsky}, {Marsden}, {Marriage},
  {Moodley}, {Niemack}, {Nolta}, {Page}, {Reese}, {Sehgal}, {Sievers},
                                                             {Spergel}, {Staggs}, \& {Wollack}}]{MenanteauET2012}
 {Menanteau}, F., {et~al.} 2012, \apj, 748, 7



\bibitem[Merten et al.(2011)]{MertenET2011}
 Merten, J., Coe, D., Dupke, R., et al.\ 2011, \mnras, 417, 333 



\bibitem[Mitchell et al.(2009)]{MitchellET2009}
 Mitchell, N.~L., McCarthy, I.~G., Bower, R.~G., Theuns, T., \& Crain, R.~A.\ 2009, \mnras, 395, 180 




\bibitem[Molnar(2015)]{Molnar2015}
 Molnar, S.~M., 2015, {\it Cosmology with Clusters of Galaxies}, 
 Nova Science Publishers, New York
 

\bibitem[Molnar(2016)]{Molnar2015Frontiers}
 Molnar, S.\ 2016, Front. Astron. Space Sci., 2, 7

\bibitem[Molnar \& Broadhurst(2017)]{MolnarBroadhurst2017}
 Molnar, S.~M., \& Broadhurst, T.\ 2017, \apj, 841, 46 


\bibitem[Molnar \& Broadhurst(2015)]{MolnarBroadhurst2015}
 Molnar, S.~M., \& Broadhurst, T.\ 2015, \apj, 800, 37 


\bibitem[Molnar et al.(2013b)]{Molnar2013ApJ774}
 Molnar, S.~M., Broadhurst, T., Umetsu, K., et al.\ 2013, \apj, 774, 70 






\bibitem[Molnar et al.(2013a)]{Molnar2013ApJ779}
 Molnar, S.~M., Chiu, I.-N.~T., Broadhurst, T., \& Stadel, J.~G.\ 2013, \apj, 779, 63 


\bibitem[Molnar et al.(2012)]{MolnarET2012ApJ748}
 Molnar, S.~M., Hearn, N.~C., \& Stadel, J.~G.\ 2012, \apj, 748, 45 




\bibitem[Molnar et al.(2010)]{Molnet10ApJ723p1272}
 Molnar, S.~M., et al.\ 2010, \apj, 723, 1272 



\bibitem[Mroczkowski et al.(2012)]{MroczkowskiET2012}
 Mroczkowski, T., Dicker, S., Sayers, J., et al.\ 2012, \apj, 761, 47  

\bibitem[{{Okabe} {et~al.}(2011){Okabe}, {Bourdin}, {Mazzotta}, \& {Maurogordato}}]{OkabeET2011}
{Okabe}, N., {Bourdin}, H., {Mazzotta}, P., \& {Maurogordato}, S. 2011, \apj, 741, 116



\bibitem[Navarro, Frenk \& White(1997)]{NFW1997ApJ490p493}
 Navarro, J.~F., Frenk, C.~S., \& White, S.~D.~M.\ 1997, \apj, 490, 493 





\bibitem[Ng et al.(2015)]{NgET2015}
 Ng, K.~Y., Dawson, W.~A., Wittman, D., et al.\ 2015, \mnras, 453, 1531 



\bibitem[Owers \etal(2011)]{OwersET2011}
 Owers, M.~S., Randall, S.~W., Nulsen, P.~E.~J., \etal, 2011, \apj, 728, 27 




\bibitem[{{Ragozzine} {et~al.}(2012){Ragozzine}, {Clowe}, {Markevitch}, {Gonzalez}, \& {Brada{\v c}}}]
               {RagozzineET2011}
 {Ragozzine}, B., {Clowe}, D., {Markevitch}, M., {Gonzalez}, A.~H., \& {Brada{\v c}}, M. 2012, \apj, 744, 94




\bibitem[Ricker(2008)]{Ricker2008ApJS176}
 Ricker, P.~M.\ 2008, \apjs, 176, 293 



\bibitem[Sayers et al.(2013)]{SayersET2013}
 Sayers, J., Mroczkowski, T., Zemcov, M., et al.\ 2013, \apj, 778, 52 


\bibitem[Springel \& Farrar(2007)]{SpringelFarrar2007MNRAS380}
 Springel, V., \& Farrar, G.~R.\ 2007, \mnras, 380, 911 


\bibitem[Stroe et al.(2014)]{StroeET2014MNRAS445}
 Stroe, A., Harwood, J.~J., Hardcastle, M.~J., \& R{\"o}ttgering, H.~J.~A.\ 2014, \mnras, 445, 1213 


\bibitem[Thompson et al.(2015)]{ThompsonET2015}
 Thompson, R., Dav{\'e}, R., \& Nagamine, K.\ 2015, \mnras, 452, 3030 



\bibitem[Thompson \& Nagamine(2012)]{ThompsonNagamine2012}
 Thompson, R., \& Nagamine, K.\ 2012, \mnras, 419, 3560




\bibitem[van Weeren et al.(2011)]{WeerenET2011MNRAS418}
 van Weeren, R.~J., Br{\"u}ggen, M., R{\"o}ttgering, H.~J.~A., \& Hoeft, M.\ 2011, \mnras, 418, 230 






\bibitem[van Weeren et al.(2011)]{WeerenET2011AA528} 
van Weeren, R.~J., Hoeft, M., R{\"o}ttgering, H.~J.~A., et al.\ 2011, \aap, 528, A38 




\bibitem[van Weeren et al.(2010)]{WeerenET2010Sci}
 van Weeren, R.~J., R{\"o}ttgering, H.~J.~A., Br{\"u}ggen, M., \& Hoeft, M.\ 2010, Science, 330, 347 



\bibitem[Watson et al.(2014)]{WatsonET2014}
 Watson, W.~A., Iliev, I.~T., Diego, J.~M., et al.\ 2014, \mnras, 437, 3776 



\bibitem[Zhang et al.(2015)]{ZhangET2015ApJ813}
 Zhang, C., Yu, Q., \& Lu, Y.\ 2015, \apj, 813, 129 


\bibitem[Zitrin et al.(2014)]{Zitrin2014ApJ793}
 Zitrin, A., Zheng, W., Broadhurst, T., et al.\ 2014, \apjl, 793, L12 




\end{thebibliography}


\end{document}